\documentclass[12pt,preprint]{aastex}

\usepackage{subfigure}
\begin{document}

\title{Comparative Lyot Coronagraphy with Extreme AO Systems}

\slugcomment{Accepted to ApJ}

\author{Justin R. Crepp, Andrew D. Vanden Heuvel, \& Jian Ge}
\affil{Astronomy Department, University of Florida \\
211 Bryant Space Science Center, P.O. Box 112055 \\
Gainesville, FL 32611-2055} \email{jcrepp@astro.ufl.edu,
avheuv@gmail.com, jge@astro.ufl.edu}

\begin{abstract}
As adaptive optics technology continues to improve, the stellar
coronagraph will play an ever increasing role in ground-based
high-contrast imaging observations. Though several different image
masks exist for the most common type of coronagraph, the Lyot
coronagraph, it is not yet clear what level of wavefront correction
must be reached in order to gain, either in starlight suppression or
observing efficiency, by implementing a more sophisticated design.
In this paper, we model image plane Lyot-style coronagraphs and test
their response to a range of wavefront correction levels, in order
to identify regimes of atmospheric compensation where the use of
hard-edge, Gaussian, and band-limited image masks becomes
observationally advantageous. To delineate performances, we
calculate the speckle noise floor mean intensity. We find that
apodized masks provide little improvement over hard-edge masks until
on-sky Strehl ratios exceed $\sim0.88 \: S_{qs}$, where $S_{qs}$ is
the intrinsic Strehl ratio provided by the optical system. Above
this value, 4th-order band-limited masks out-perform Gaussian masks
by generating comparable contrast with higher Lyot stop throughput.
Below this level of correction, hard-edge masks may be
preferentially chosen, since they are less susceptible to low-order
aberration content. The use of higher-order band-limited masks is
relegated to situations where quasi-static residual starlight cannot
be sufficiently removed from the search area with speckle-nulling
hardware.
\end{abstract}






\section{INTRODUCTION}
In recent years, great strides in the development of adaptive optics
(AO) technology have ushered in a new era of high resolution
diffraction-limited imaging. Despite these advances, the ability of
the stellar coronagraph to generate deep contrast remains limited by
insufficient levels of wavefront control; uncorrected phase and
amplitude errors induced by the atmosphere and instrument optics
manifest as bright, dynamic `speckles' of scattered light in the
search area. Even the most basic coronagraph, a Lyot coronagraph
equipped with a focal plane hard-edge occulter (Lyot 1939), is
incapable of reaching its peak performance when coupled to
state-of-the-art AO (Oppenheimer 2004).

Numerous high-contrast observations have been conducted using AO on
the world's largest telescopes (Marois et al. 2006, Mayama et al.
2006, Itoh, Oasa, \& Fukagawa 2006, Carson et al. 2005a, Close et
al. 2005, Metchev \& Hillenbrand 2004, Debes et al. 2002, Liu et al.
2002, and references therein); some rely solely on AO imaging, while
others combine AO with coronagraphy\footnote{Interferometers are
likewise capable of suppressing starlight, and generally have a
better inner-working-angle but a more restricted search area (Absil
et al. 2006, Serabyn et al. 2005).} and/or speckle reduction
techniques. These efforts have provided the first image of a
candidate extrasolar planet (Chauvin et al. 2004), as well as direct
detections of sub-stellar companions near the planet-brown dwarf
boundary (Biller et al. 2006, Neuh{\"a}user et al. 2005, Chauvin et
al. 2005). However, to image older, less-massive, and closer-in
planets from the ground, wavefront sensing and correction techniques
must improve substantially.

``Extreme'' advances in high-contrast imaging technology are
anticipated in the coming years. Deformable mirrors employing
several thousand actuators and wavefront sensing of laser guide
stars can, in principle, drive Strehl ratios above $90\%$ on 8-10m
class telescopes. With the proper coronagraph, these systems will be
capable of detecting the near-IR emission of Jovian planets over a
broader range of ages, masses, and separations (see Macintosh et al.
2003). Extremely large telescopes such as the proposed Thirty Meter
Telescope (TMT; Macintosh et al. 2006, Troy et al. 2006, Ellerbroek
et al. 2005) and 100m OverWhelmingly Large telescope (OWL; Brunetto
et al. 2004) will improve spatial resolution, and hence the
inner-working-angle (hereafter IWA) on the sky.

An interesting and more immediate alternative, which uses current AO
technology, can also provide highly corrected wavefronts by instead
sacrificing spatial resolution in return for improved pupil
sampling. Serabyn et al. 2007 have demonstrated K-band Strehl ratios
approaching $94\%$ by reimaging a 1.5m diameter circular
unobstructed subaperture of the Hale 200-inch telescope onto the
existing deformable mirror via relay optics. Combining this
technique with a coronagraph that has an intrinsically small IWA ($<
3 \, \lambda / D$) shows promise for generating deep contrast for
separations as close as $\sim 0.5$".

These considerations motivate the need for a quantitative
understanding of how the stellar coronagraph's utility will depend
on future gains in AO proficiency. To address this topic, we have
modeled systems that are equipped with a variety of amplitude image
masks, and examined their performance in a broad, nearly continuous
range of corrected wavefront levels. With this paper, we seek to
provide a concise guide to the use of Lyot-style image plane
coronagraphs. Most notably, we answer the question: ``What is the
appropriate choice of image mask for an extreme AO system operating
at a given Strehl ratio?''.

Before proceeding however it is important to discuss an impediment
that currently burdens all of the observational surveys listed
above. In the next section, we describe quasi-static aberrations and
the potential to suppress them to levels sufficient for direct
substellar companion detection. The topical outline for the
remainder of the paper consists of our numerical techniques,
results, and summary and concluding remarks, which are discussed in
$\S\ref{sec:numerical}$, $\S\ref{sec:comparative}$, and
$\S\ref{sec:conclusions}$ respectively.


\section{Quasi-Static Aberrations}
\label{sec:qs_aberr}

In addition to atmospheric turbulence, perturbations to the optical
path likewise occur once starlight reaches the telescope and
instrument optics. Of considerable concern are errors introduced
downstream from the AO deformable mirror (DM) and wavefront sensor.
The effects of these more systematic disturbances manifest as slowly
undulating, quasi-static speckles in the image plane that change on
a timescale commensurate with the telescope tracking, thermal
fluctuations, and/or other flexures to the opto-mechanical system.
Such aberrations are inherent to all instruments and unavoidable at
the contrast levels required for circumstellar science.



Speckle-nulling is the act of removing quasi-static residual
starlight from a pre-selected region of the image plane. By
including one or more additional DMs in the optical train pupil
planes to reshape the phase of starlight, a sharply defined region
of deeper contrast can be generated over a fraction of the search
area. The size of this so-called ``dark hole" is governed by the
number of extra DMs and their actuator densities. This technique
greatly improves the chances for direct detection by properly
isolating the companion's mutually incoherent signal. Laboratory
results from the High Contrast Imaging Testbed (HCIT; Trauger et al.
2004, Borde et al. 2006) have shown that speckle-nulling is indeed
quite feasible down to the contrast levels potentially accessible to
ground-based coronagraphs. High-contrast imaging systems should be
equipped with such capabilities, since explicit removal of scattered
light with hardware will yield the most unambiguous sub-stellar
companion detections and spectra.

Differential imaging techniques such as Simultaneous Differential
Imaging (SDI; Racine et al. 1999, Marois et al. 2005, Close et al.
2005, Biller et al. 2006), Angular Differential Imaging (ADI; Marois
et al. 2006), and others (see Guyon 2004 and references within) can
also help to circumvent the problem of quasi-static aberrations. For
example, by subtracting images taken simultaneously in different,
carefully chosen bandpasses (SDI) or subsequently at different
roll-angles (ADI), it is possible to enhance the effective contrast
and detect companions whose brightness falls below the speckle noise
floor. These techniques are quite powerful and can even be used in
concert. They are limited only by the standard deviation of the
stellar residual signal.

In the following, we compare image mask performances by calculating
the direct output of the coronagraph, i.e. the mean intensity, with
the understanding that differential and post-processing techniques
can always be used on top of direct imaging to improve the prospects
for discovery.

\section{NUMERICAL SIMULATIONS}
\label{sec:numerical}

The theoretical model used for our simulations consists of an
extreme AO system linked in series to a Lyot coronagraph that is
observing a stellar point source. Wavefront correction levels
spanning from not quite diffraction limited ($\sim 77\%$ Strehl) to
highly corrected ($\sim 96\%$ Strehl) are generated with an IDL
routine based on simulations described in Carson et al. 2005b. The
code is optimized to simulate the operation of PHARO (Hayward et al.
2001) with the PALAO system (Troy et al. 2000) on the Hale 200-inch
telescope at Palomar.


To clearly elucidate the sometimes subtle differences in performance
between coronagraphic image masks, we restrict our analysis to
monochromatic light ($\lambda = 2.2 \, \mu m$), and ignore the
effects of central obstructions, their support structures, and
inter-segment mirror gaps. To first order, the addition of each of
these complexities can be understood by convolving the telescope
entrance aperture with the spatial frequency spectrum of the mask,
and observing the resultant light distribution in the Lyot plane.
The net effect is often simply a loss in off-axis throughput, as the
Lyot stop size is necessarily reduced to reject the additional
diffracted starlight. Abe et al. 2006, Sivaramakrishnan \& Yaitskova
2005, Sivaramakrishnan \& Lloyd 2005, Soummer 2005, and Murakami \&
Baba 2005 discuss the prospects for Lyot coronagraphy with
non-trivial entrance aperture geometries. We use a circular
unobstructed entrance aperture, radial image mask, and (hence) a
circular Lyot stop.


Kolmogorov phase screens mimic the effects of atmospheric
turbulence, where a fixed Fried parameter of 20 cm at 2.0 microns,
which has previously been found to best match actual PHARO data
(Carson et al. 2005b), is used throughout. To emulate AO correction,
the phase screens are Fourier transformed, multiplied by a parabolic
high-pass filter (Sivaramakrishnan et al 2001, Makidon et al. 2005),
and then inverse-transformed. Improving the degree of wavefront
correction is accomplished by increasing the actuator density,
which, in turn, raises the critical frequency of the high-pass
spatial filter. The linear number of actuators across the pupil
ranges from 35 to 94 (962 to 6939 total actuators); in terms of
root-mean-square (rms) residual error, this provides a range in
correction from $\lambda/13$ to $\lambda/30$.


The resulting AO-corrected wavefronts are then sent to a separate
MATLAB coronagraph code for analysis, where the starlight passes
through a series of consecutive pupil and image planes. We assume
idealized interactions with the optical elements and the image mask
(i.e. no scattered light, dust, fabrication errors, ... etc.), and
use scalar diffraction theory to calculate the propagation of the
electric field. The telescope pupil is constructed as a perfect disk
of unit transmission and diameter 512 pixels placed at the center of
a 3072 x 3072 padding matrix in order to provide sufficient image
plane resolution (6 pixels per $\lambda/D$). This choice of matrice
sizes results in numerical noise levels below $10^{-12}$ in
contrast, which is negligible compared to the physical speckle noise
floor set by the AO system.


We do not explicitly simulate speckle-nulling on top of atmospheric
correction. Instead, in $\S$\ref{sec:hard_vs_apo} and
$\S$\ref{sec:gauss_vs_bl}, we assert that ``the instrument" corrects
for quasi-static aberrations down to the noise floor mean intensity
set by diffraction and the atmosphere, for a given AO actuator
density. Speckle-nulling is discussed in more depth in
$\S$\ref{sec:tip_tilt}. This assumption is justified given the
laboratory experiments mentioned in $\S$\ref{sec:qs_aberr} that
demonstrate removal of residual starlight to contrast levels below
$10^{-9}$ within the fractional search area; although, in practice
more timely methods for finding the optimal shapes of the extra DMs
(preferably $<$ 1 minute) will need to be employed.













\section{COMPARATIVE LYOT CORONAGRAPHY}
\label{sec:comparative}

Our study focuses on the subclass of Lyot coronagraphs that control
diffracted starlight with amplitude image masks - that is,
focal-plane masks that do not modulate the phase of transmitted
light in theory.\footnote{Masks that manipulate the phase of
starlight in the image plane generally have better
inner-working-angles but poorer broadband performance (Roddier \&
Roddier 1997 and Rouan et al. 2000); although, fully achromatic
designs are being developed (e.g. Mawet et al. 2005). See Guyon et
al. 2006 for a review of the myriad of other different coronagraphic
designs and how they compare in a space-based application, such as
TPF.} Such masks reside in the focal-plane wheel of many
coronagraphs in operation at major observatories. Among the choices
of amplitude image masks, band-limited masks (Kuchner \& Traub 2002)
can perform the best in principle. In the ideal case, they diffract
{\it all} transmitted on-axis starlight into a narrow region
surrounding the edges of the Lyot pupil, and leave an area of
infinite dynamic range in the center. Moreover, band-limited masks
with arbitrarily broad central nulls can also be constructed
(Kuchner, Crepp, \& Ge 2005), and have been shown to help combat
low-spatial-frequency optical aberrations in numerical simulations
(Shaklan \& Green 2005) and in laboratory experiments (Crepp et al.
2006).


We include a hard-edge mask, a Gaussian mask, and band-limited
(hereafter BL) masks with 4th-order ($\mbox{sinc}^2$), 8th-order
($\mbox{sinc}$, $\mbox{sinc}^3$), and 12th-order ($\mbox{sinc}$,
$\mbox{sinc}^2$, $\mbox{sinc}^3$) intensity transmission profiles
near the optical axis (Figure 1). Each mask is azimuthally symmetric
and designed with an IWA$\, =\, 4\; \lambda/D$, so as to make fair
comparisons. The IWA is defined as the half-width-at-half-maximum of
the intensity transmission profile; for these masks, this value
differs by less than $1\%$ from the equivalent width, which can also
be used as an alternative definition (Aime 2005). The masks are not
truncated in the image plane; in practice, it is easy to include
enough resolution elements such that this effect does not contribute
significantly to the noise floor. Equations describing the masks are
shown below. The radial coordinate, $r$, measures the distance from
the optical axis, where $\tilde r = r \; D / \lambda$. Constants,
which can be derived from Kuchner, Crepp, \& Ge 2005, are given to
four decimals of precision. The amplitude transmissions, $M(r)$,
are:
\begin{eqnarray}
M_{\mbox{H}}(r)&=& \mbox{circ} \: (\tilde r / 4) \\ 
M_{\mbox{G}}(r)&=& 1 - e^{-(\tilde r / 3.6097)^2}, \label{equ:gaussian} \\
M_{{\mbox{BL}_{4th}}}(r)&=& 1 - \mbox{sinc}^2(0.4500 \; \tilde r),\\
M_{{\mbox{BL}_{8th}}}(r)&=& 0.9485+0.4743 \; \mbox{sinc}(1.4043 \; \tilde r)\\
\nonumber             &&  - \; 1.4228 \; \mbox{sinc}^3(0.4681 \; \tilde r),\\
M_{{\mbox{BL}_{12th}}}(r)&=& 0.7526 - 0.9408\;\mbox{sinc}(1.9006\; \tilde r) \\
\nonumber             &&  + \; 5.2684 \; \mbox{sinc}^2(0.9503\; \tilde r) \\
\nonumber             &&  - \; 5.0802 \; \mbox{sinc}^3(0.6335\;
\tilde r),
\end{eqnarray}
where $\mbox{circ} \: (\tilde r/a)$ is a step-function equal to zero
for $\tilde r \leq a$ and unity elsewhere.

To provide intuition for each mask's potential performance, we first
present coronagraph simulations using perfect incident wavefronts
(Table~\ref{tab:comparison}). A qualitative understanding of the
coronagraph's functionality can be gleaned by examining the light
distribution pattern in the Lyot pupil plane, since the total amount
of rejected starlight depends strongly on the Lyot stop size and
shape. For hard-edge and Gaussian masks, the contrast is limited by
residual diffracted starlight, whereas the combination of a BL mask
with a matching Lyot stop can completely remove all on-axis
starlight, in this ideal case. This capability is seen in the 4th-,
8th-, and 12th-order BL mask final image plane patterns; they are
composed entirely of numerical noise.

``Degeneracies'' in contrast can occur however when large phase
errors are present. In other words, different masks may generate
indistinguishably similar levels of contrast when uncorrected
atmospheric scattered starlight, rather than diffracted starlight,
is the dominant source of noise. Under these circumstances,
throughput, quasi-static aberration sensitivities, and fabrication
considerations should more heavily influence the decision for which
mask to implement in practice.


In the following sections, image mask performances are quantified in
terms of these parameters and as a function of AO system correction.
References to both the rms wavefront error (WFE), $\sigma_{AO}$, in
the pupil plane and resulting Strehl ratio, $S_{AO}$, in the image
plane are made. The relationship $S_{AO} \approx 1-(2 \pi
\sigma_{AO} )^2$ is valid in the high Strehl regime, and is adopted
for our calculations. We begin by comparing hard-edge masks to the
class of graded or apodized masks as a whole.

At this point it is convenient to define contrast, $C(r)$, as it
will be used throughout the remainder of the text:
\begin{equation}
C(r)=\frac{I(r)}{\hat I(0) \, |M(r)|^2},
\end{equation}
where $I(r)$ is the intensity at the radial coordinate in the final
image, $\hat I(0)$ is the peak stellar intensity as would be
measured {\it without} the image mask in the optical train, and
$|M(r)|^2$ is the mask intensity transmission. Note that $I(r)$ and
$\hat I(0)$ are both measured with the Lyot stop in place. This is
the non-differential (i.e non-SDI, single roll-angle, ... etc.)
contrast.



\subsection{Hard-Edge vs. Apodized Image Masks}
\label{sec:hard_vs_apo}

A stellar coronagraph will begin to noticeably improve contrast
relative to standard AO imaging only when a sufficient amount of
starlight is concentrated in the Airy pattern central core. This
occurs at Strehl ratios as low as $50\%$; however, substantial gains
over a significant fraction of the search area are not attained
until the Strehl ratio exceeds $\sim 80\%$ (Sivaramakrishnan et al.
2001). We seek to identify the required level of wavefront
correction above this value where apodized masks begin to
out-perform hard-edge masks.\footnote{Notice that the IWA is rather
loosely defined, and apodized masks can, in principle, work interior
to the hard-edge mask. This benefit is likely difficult to exploit
in practice, but constitutes an important caveat to the analysis. We
assume that speckle-nulling is only applied exterior to the IWA.}

For this comparison, we hold the Lyot stop size fixed at 60\%
throughput; images from the Lyot Plane column in
Table~\ref{tab:comparison} and results from
$\S\ref{sec:gauss_vs_bl}$ help to clarify why this is a useful
simulation control. Contrast curves for the hard-edge, Gaussian, and
4th-order BL masks are shown in Figure~\ref{fig:hard_apo} using
several different levels of wavefront correction. For clarity,
higher-order BL masks are not included here, but are discussed in
the next sections. They provide intermediate levels of contrast,
between that of the hard-edge and the 4th-order BL mask.

At $\sim 77\%$ Strehl (rms WFE $= \lambda / 13$), the advantage in
using the image plane coronagraph is evident only for separations
smaller than $\sim 8 \; \lambda / D$; this is a result of the
limited wavefront correction and reduced aperture stop in the Lyot
plane. Moreover, the Gaussian and 4th-order BL masks provide only a
factor of $\sim 1.6$ improvement over the hard-edge occulter at the
IWA, and contrast degeneracy sets in at a distance of $\sim 1.5 \,
\lambda / D$ from there. As AO correction improves, starlight is
redistributed from the scattered light halo into the Airy
diffraction pattern, the PSF `shoulder' drops and extends, and the
diffracted light limitations of the hard-edge mask are revealed. The
sharp edges of the mask prevent further improvements in contrast
below $\sim 10^{-4}$ at the IWA, even as Strehl ratios exceed $\sim
88\%$.


The Gaussian and 4th-order BL masks, which more effectively diffract
starlight onto the opaque portions of the Lyot stop, are capable of
generating contrast levels of $\sim 6 \times 10^{-6}$ at $\mbox{IWA}
= 4 \; \lambda / D$ with $\sim 94\%$ Strehl. They provide an
improvement in contrast at the IWA over the hard-edge mask by
factors of approximately 2.7 at $\sim 82\%$ Strehl (rms WFE $=
\lambda / 15$), 5.1 at $\sim 88\%$ Strehl (rms WFE $= \lambda /
18$), and 21.7 at $\sim 94\%$ Strehl (rms WFE $= \lambda / 26$).
These results are consistent with the hard-edge versus Gaussian mask
lab experiments of Park et al. 2006, taking into consideration the
differences in PSF structure and IWAs.

Curves showing contrast at the IWA and the relative improvement are
shown in Figure~\ref{fig:continuous_hard_apo} as a more continuous
function of wavefront correction. Though Gaussian and 4th-order BL
masks out-perform the hard-edge occulter at currently achievable
Strehl ratios, they remain degenerate with one another into the
realm of extreme AO. We next identify the conditions and parameters
for which this performance degeneracy is broken.

\subsection{Gaussian vs. Band-Limited Masks}
\label{sec:gauss_vs_bl}

In order to compare apodized masks to one another, consideration of
the light distribution pattern in the Lyot pupil plane must be made.
It is clear from the previous section that Gaussian masks are
competitive with BL masks in terms of contrast even at very high
Strehl ratios. Thus, we can use the Lyot plane patterns shown in the
third column of Table~\ref{tab:comparison} for qualitative guidance.

Consider the effect of changing the Lyot stop diameter, $D_L$, for
each of the apodized masks. With small non-zero stop sizes, the
contrast at a given location on the detector is governed by the
competing effects of speckle noise intensity and the achievable peak
intensity of an off-axis source. The speckle noise intensity scales
as $D_L^2$, whereas the companion peak intensity, $\hat I(0) \:
|M(r)|^2$, scales as $D_L^4$. The result is a net improvement in
contrast and the apodized masks perform comparably, until $D_L$
grows large enough to leak significant levels of diffracted
starlight.

As the Lyot stop size increases further, the contrast generated by
BL masks does not degrade in as smooth a fashion as the Gaussian
mask, since the diffracted light is tightly concentrated into a
small region that follows the contour of the telscope entrance
aperture. Instead, the transition away from optimal performance is
more abrupt. The width of the transition region in the Lyot plane
narrows as the wave front correction improves, and the transition
occurs at progressively smaller Lyot stop sizes for masks of
higher-order (8th-, 12th-, ... etc.).



These effects are seen in Figure~\ref{fig:ls}, where we plot
contrast at the IWA against Lyot stop throughput for two levels of
wavefront correction. At $\sim 90\%$ Strehl (rms WFE $\,=
\lambda/20$), the apodized masks are clustered to within an order of
unity in contrast from one another, but clearly out-perform the
hard-edge occulter. At $\sim 96\%$ Strehl (rms WFE $\,=
\lambda/30$), the apodized mask performance degeneracy is broken,
and optimum Lyot stop sizes become more evident. In particular, the
4th-order band-limited mask affords an $\sim 10\%$ gain in Lyot stop
throughput over the Gaussian mask ($60\%$ vs. $50\%$). The 8th-order
and 12th-order masks generate slightly worse contrast and with
$\lesssim 40\%$ and $\lesssim 15\%$ throughput respectively. These
exact values depend upon the entrance aperture geometry, operating
bandwidth, IWA, mask-function, and mask-type (linear, radial, or
separable), but the trend is nevertheless the same at this level of
wave front correction.



Generating Strehl ratios beyond 96\% on large ground-based
telescopes in the near future is rather unlikely.\footnote{The JWST
however will provide a unique and stable platform for coronagraphy
in the $3-5 \,\mu$m band from space (Green et al. 2005).} Additional
complications such as differential chromatic wavefront sensing and
correction limitations and photon noise are significant at this
level (Nakajima 2006, Guyon 2005). The potential benefits in using
higher-order BL masks from the ground are thus restricted to
guarding against low-order aberrations introduced downstream from
the AO system, but at the expense of contrast, throughput, and
angular resolution.





\subsection{Tip/Tilt \& Low-order Aberration Sensitivities}
\label{sec:tip_tilt}


Thus far we have neglected quasi-static phase errors, or, at least,
have assumed that the resulting scattered light has been judiciously
removed with speckle-nulling hardware.\footnote{Actually the light
must go somewhere in order to conserve energy. It is sometimes
preferable to simply place it on the other side of the image plane
during a given integration.} In this section we take a closer look
at the issue. To get a feel for the problem, we calculate the
contrast degradation due to just one error, systematic misalignment,
at several characteristic levels of AO correction. Then we combine
the results with theoretical low-order phase aberration information
gathered from other studies and discuss the implications. This
analysis along with that laid out in the previous two sections
provides important first-order guidelines for selecting
coronagraphic image masks for extreme AO systems.


Image mask response to quasi-static aberrations impacts the dynamic
range and duty-cycle efficiency of high-contrast observations. For
instance, in the case of tip/tilt errors, on-sky tracking latency or
telescope-to-instrument flexure may lead to misalignments that leak
significant amounts of light. Clearly, the masks presented here have
different levels of resistance to such errors.


To assess pointing sensitivities, systematic tilt phase aberrations
were added to the wavefront at the AO-coronagraph (IDL-MATLAB)
interface. Optimum Lyot stop sizes were used for each mask, and
median, instead of mean, intensities were evaluated to prevent bias
towards poor contrast. The $1 \, \lambda/D$ annulus directly outside
the coronagraph IWA remained centered on the star as it was
methodically shifted relative to the mask. Results are shown in
Fig.~\ref{fig:tt} for linear alignment errors up to $5 \, \lambda /
D$.




At relatively low Strehl ratios ($\lesssim 88\%$), hard-edge masks
perform comparably with apodized masks provided that the pointing
error does not exceed $\sim 2 \; \lambda / D$. At higher levels of
correction ($\gtrsim 88\%$), mask alignment becomes more critical.
In this regime, apodized masks are capable of generating
significantly deeper contrast than the hard-edge mask. In
particular, the Gaussian and 4th-order BL masks provide optimum
contrast when aligned to better than $\sim 1 \, \lambda / D$ at
$\sim 88\%$ Strehl and $\sim 0.5 \, \lambda / D$ at $\sim 94\%$
Strehl. If such accuracy is difficult to manage, higher-order BL
masks may be chosen over the Gaussian and 4th-order BL mask, with
the usual tradeoffs ($\S$\ref{sec:gauss_vs_bl}).



\begin{figure}[!ht]
\centering
\includegraphics[width=3in]{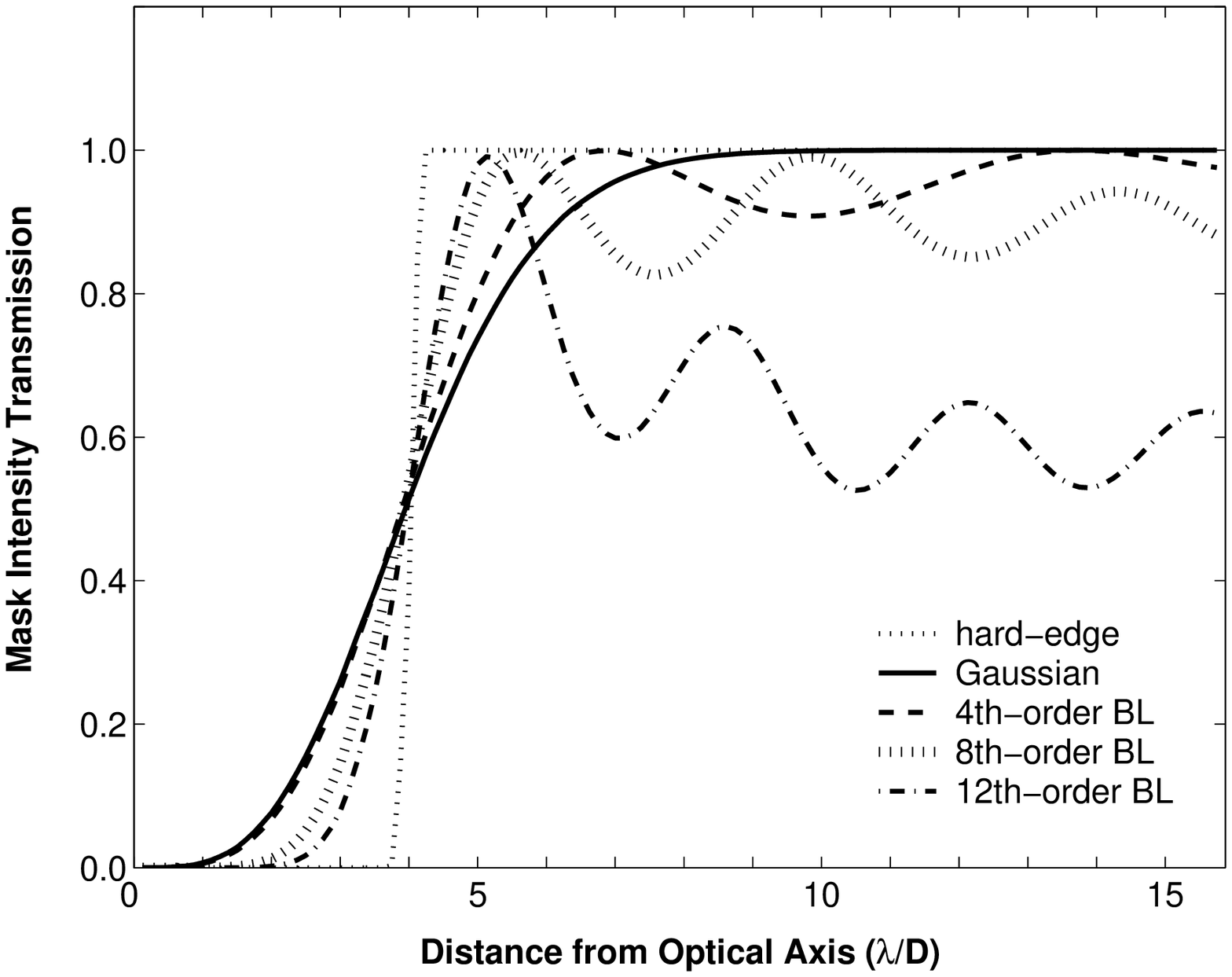}
\caption{Intensity transmission profiles for each of the radial
image masks. The IWA$\, =\, 4\; \lambda/D$. Band-limited (BL) masks
have extended off-axis attenuation, which allows them to be composed
of a finite range of low spatial-frequencies. This feature offers
unlimited dynamic range as the Strehl ratio approaches $100\%$. A
Lyot-style coronagraph equipped with a BL mask is one of the leading
candidate designs for the Terrestrial Planet Finder Coronagraph
({\it TPF-C}) space mission (Ford et al. 2004).} \label{fig:masks}
\end{figure}

\begin{figure}
\centerline{
\includegraphics[width=3.5in]{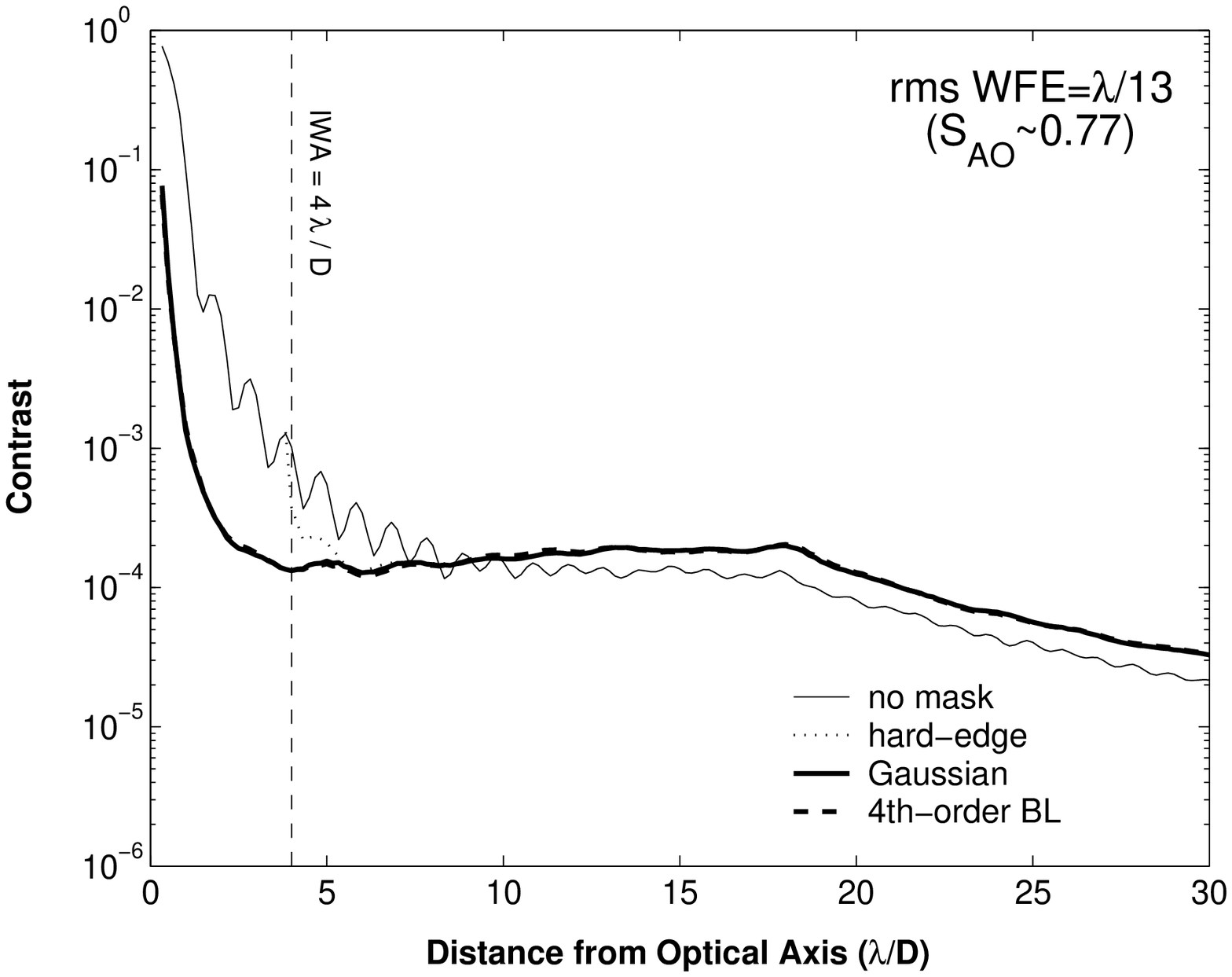}
\hfill
\includegraphics[width=3.5in]{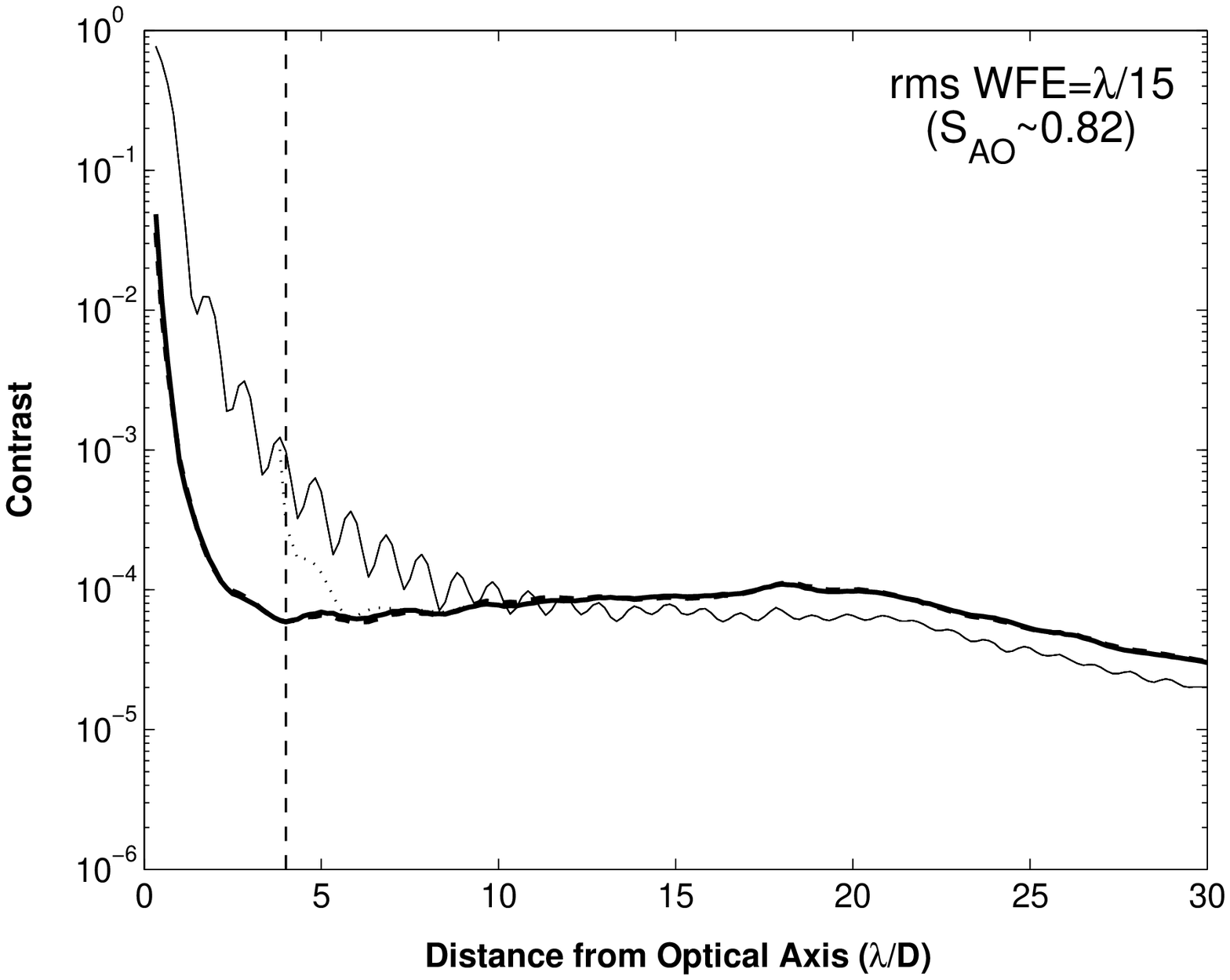}}
\centerline{
\includegraphics[width=3.5in]{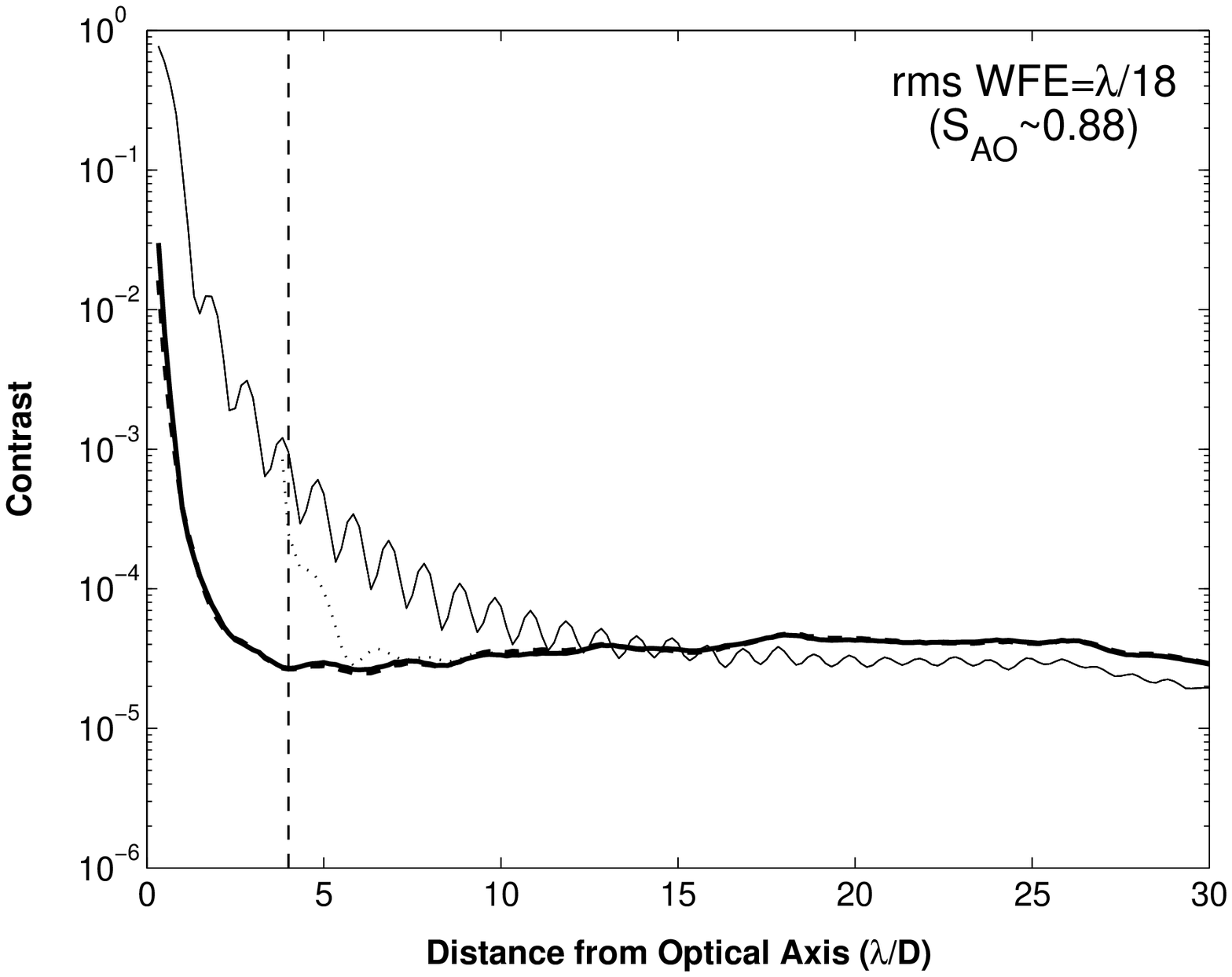}
\hfill
\includegraphics[width=3.5in]{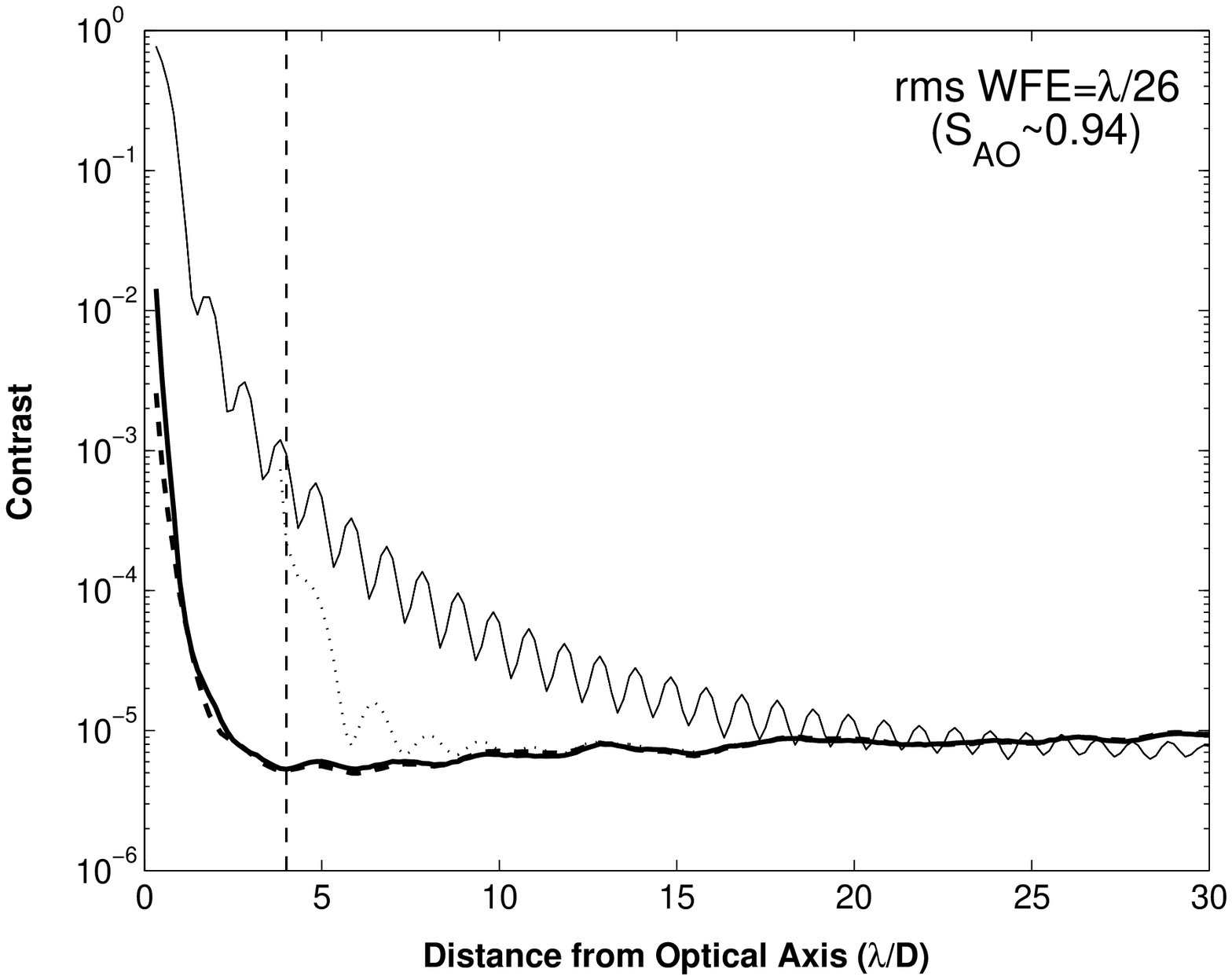}}
\caption{Azimuthally averaged contrast curves for the hard-edge,
Gaussian, and 4th-order BL image masks using a fixed circular Lyot
stop size with $\sim$60$\%$ throughput. The Lyot stop for the Airy
pattern has 100\% throughput.} \label{fig:hard_apo}
\end{figure}

\begin{figure}[!ht]
\centerline{
\includegraphics[height=3.0in]{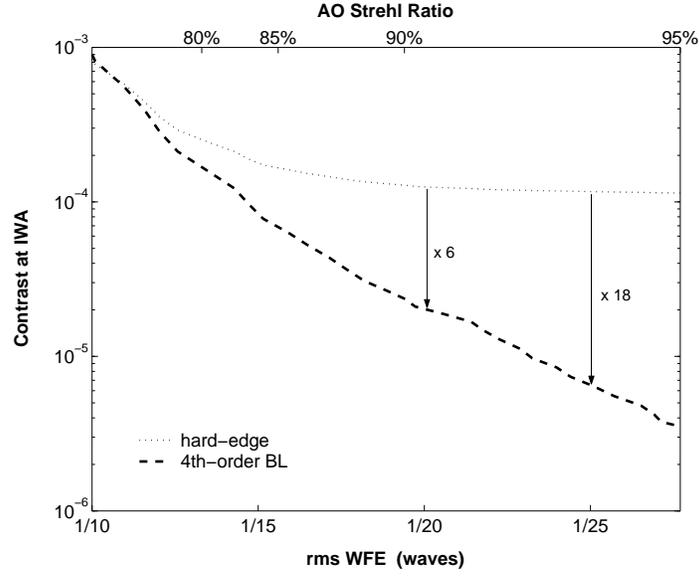}
} \caption{Average contrast within the $1 \, \lambda/D$ wide annulus
directly outside of the coronagraph IWA as a continuous function of
atmospheric wavefront correction and the relative gain (shown by
arrows) achieved by switching to the 4th-order BL mask. The
hard-edge mask's performance is limited by starlight diffracted into
the interior of the $\sim60\%$ throughput Lyot stop. This prevents
significant improvements beyond $\sim 88\%$ Strehl.}
\label{fig:continuous_hard_apo}
\end{figure}

\begin{figure}[h!t]
\centerline{
\includegraphics[height=2.5in]{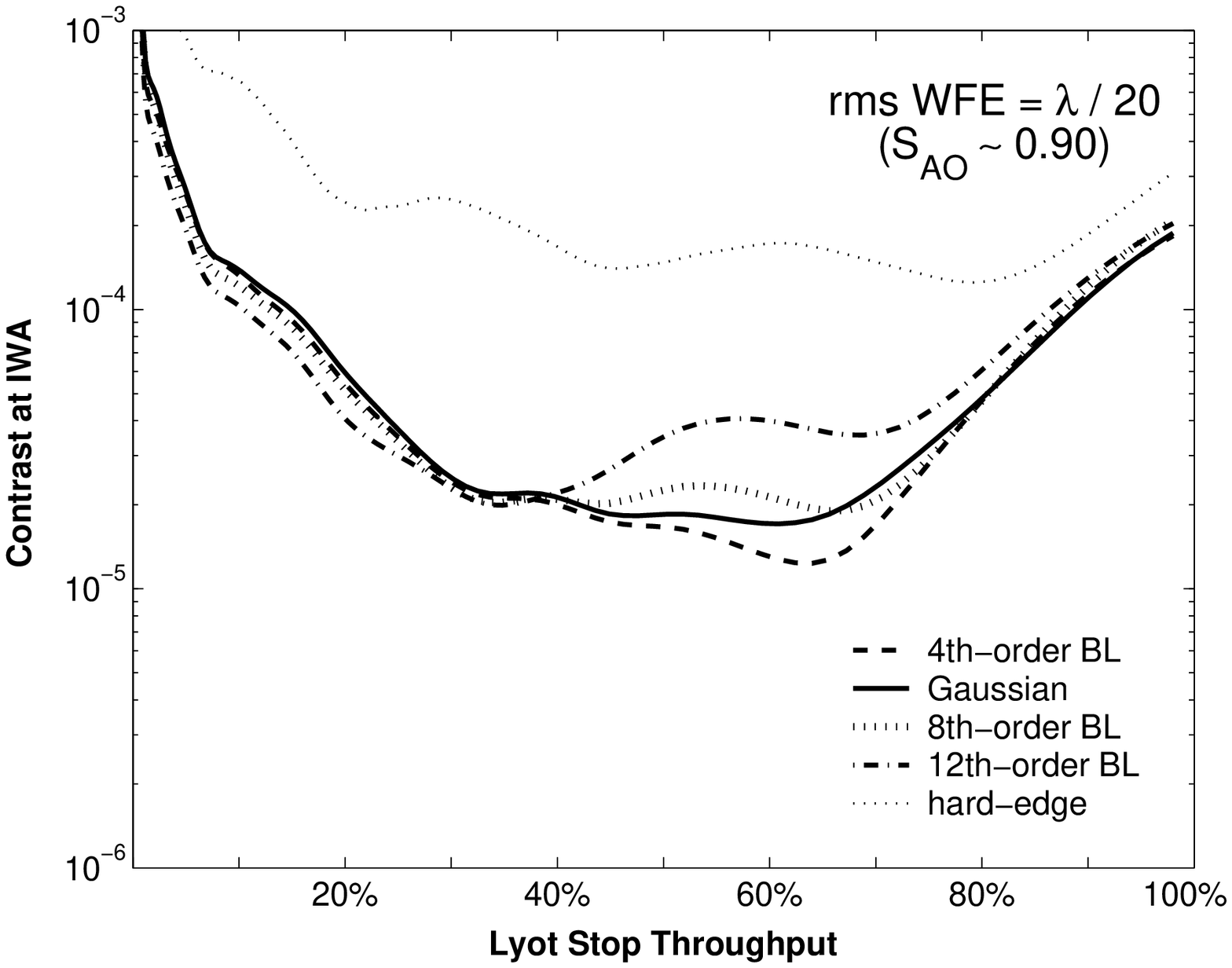}
\hfill
\includegraphics[height=2.5in]{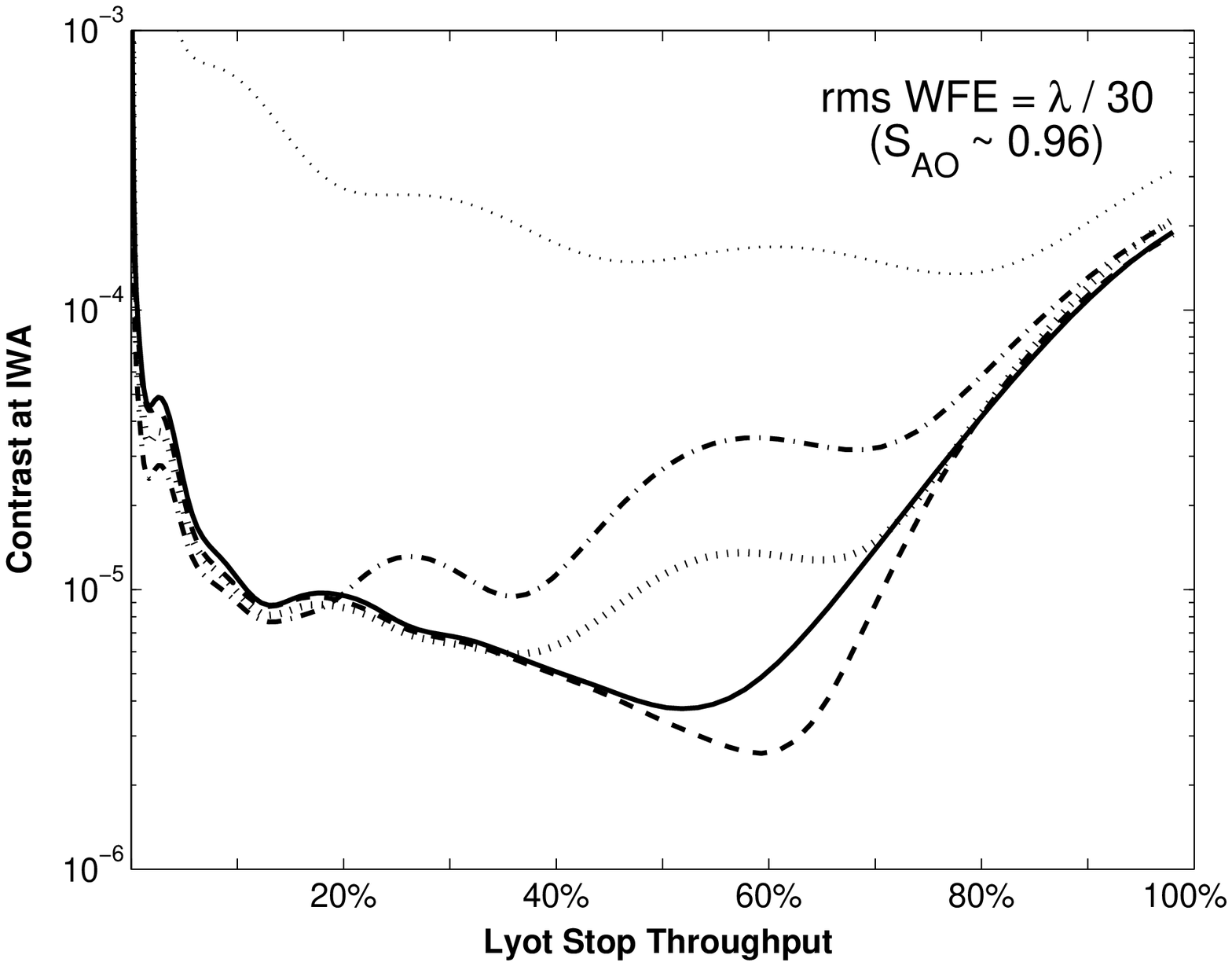}}
\caption{Average contrast within the $1 \,\lambda/D$ wide annulus
directly outside the coronagraph IWA as a function of Lyot stop
throughput for $\sim 90\%$ Strehl (left) and $\sim 96\%$ Strehl
(right).} \label{fig:ls}
\end{figure}

\begin{figure}[!h]
\centerline{
\includegraphics[width=3.1in]{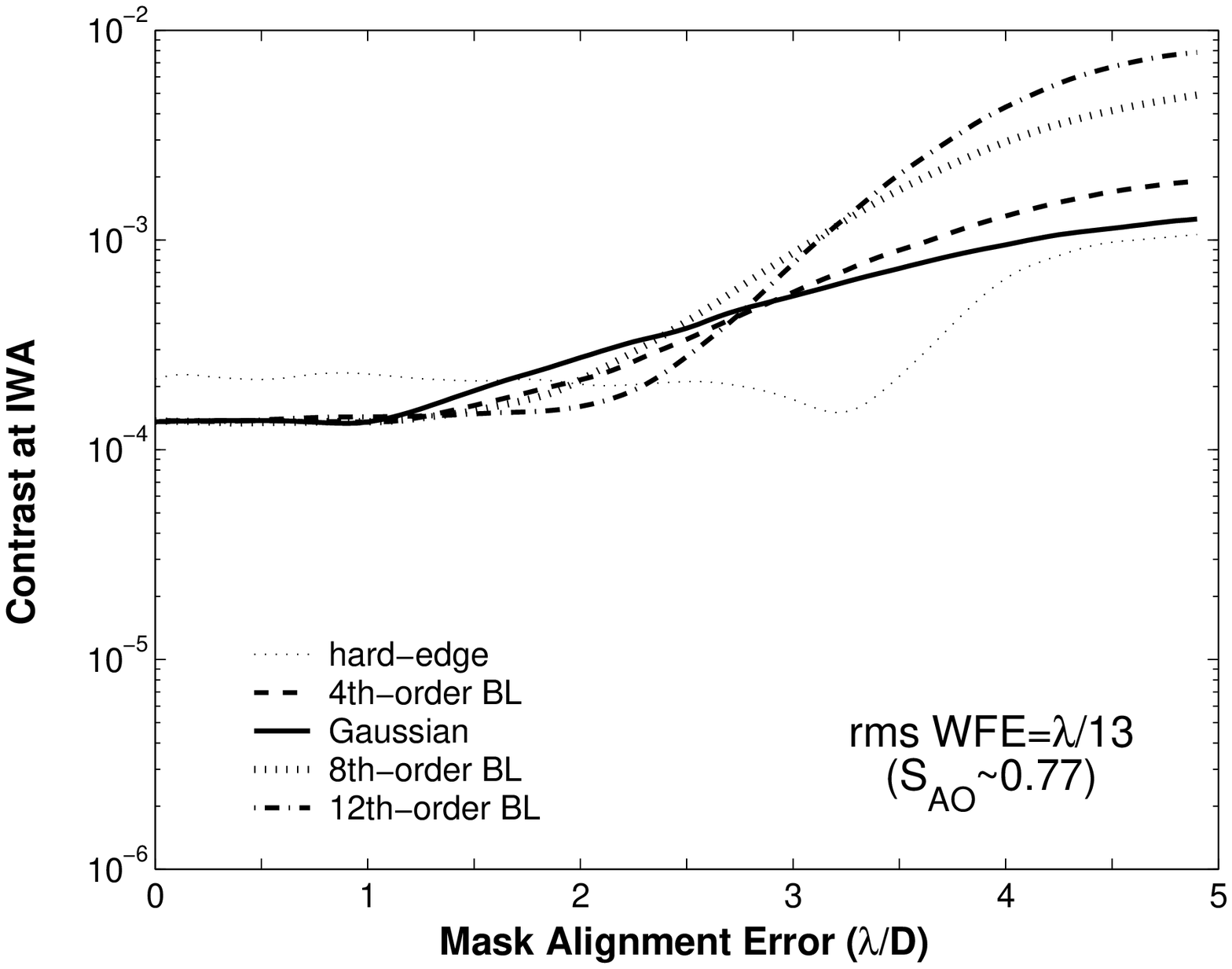}
\hfill
\includegraphics[width=3.1in]{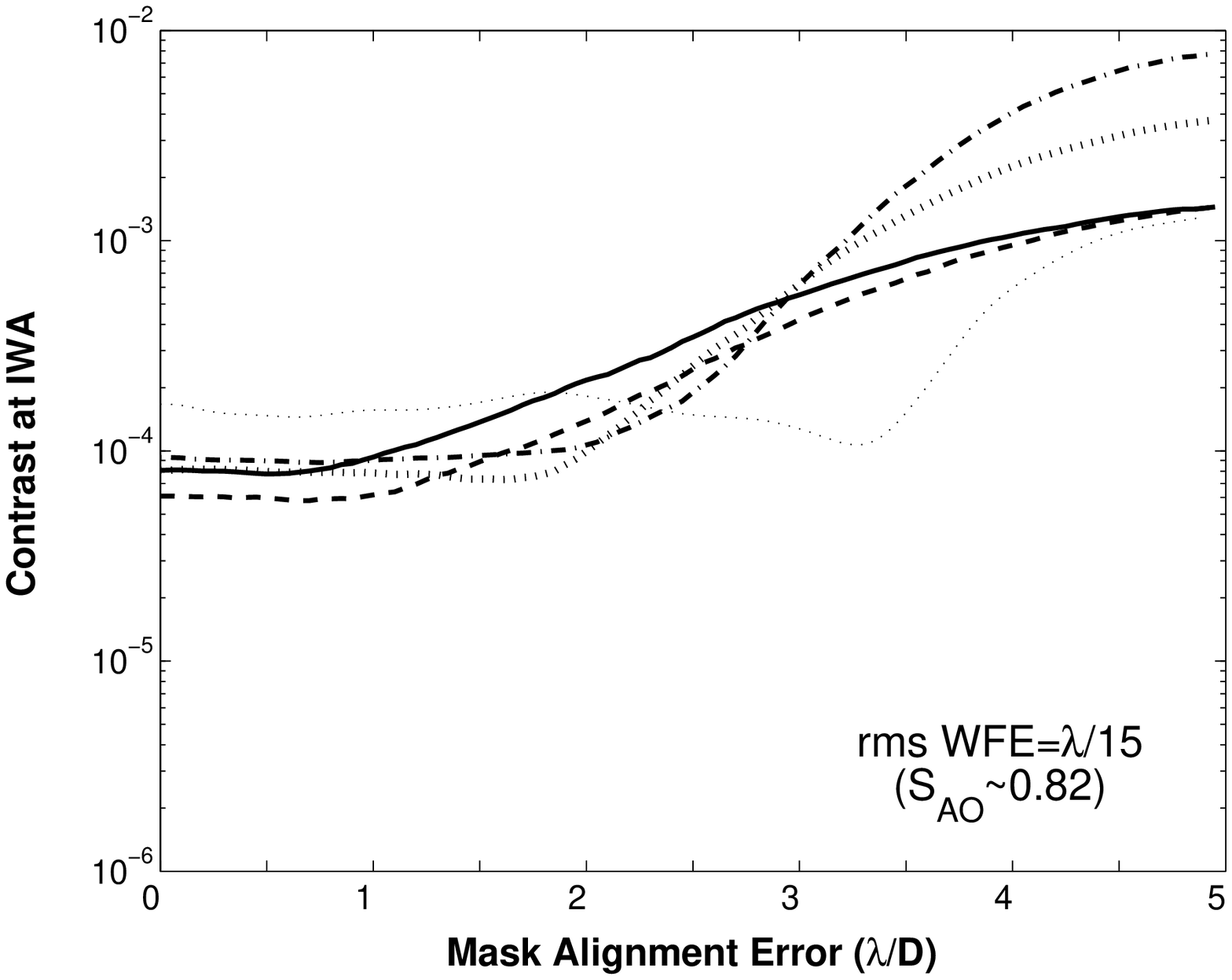}}
\centerline{
\includegraphics[width=3.1in]{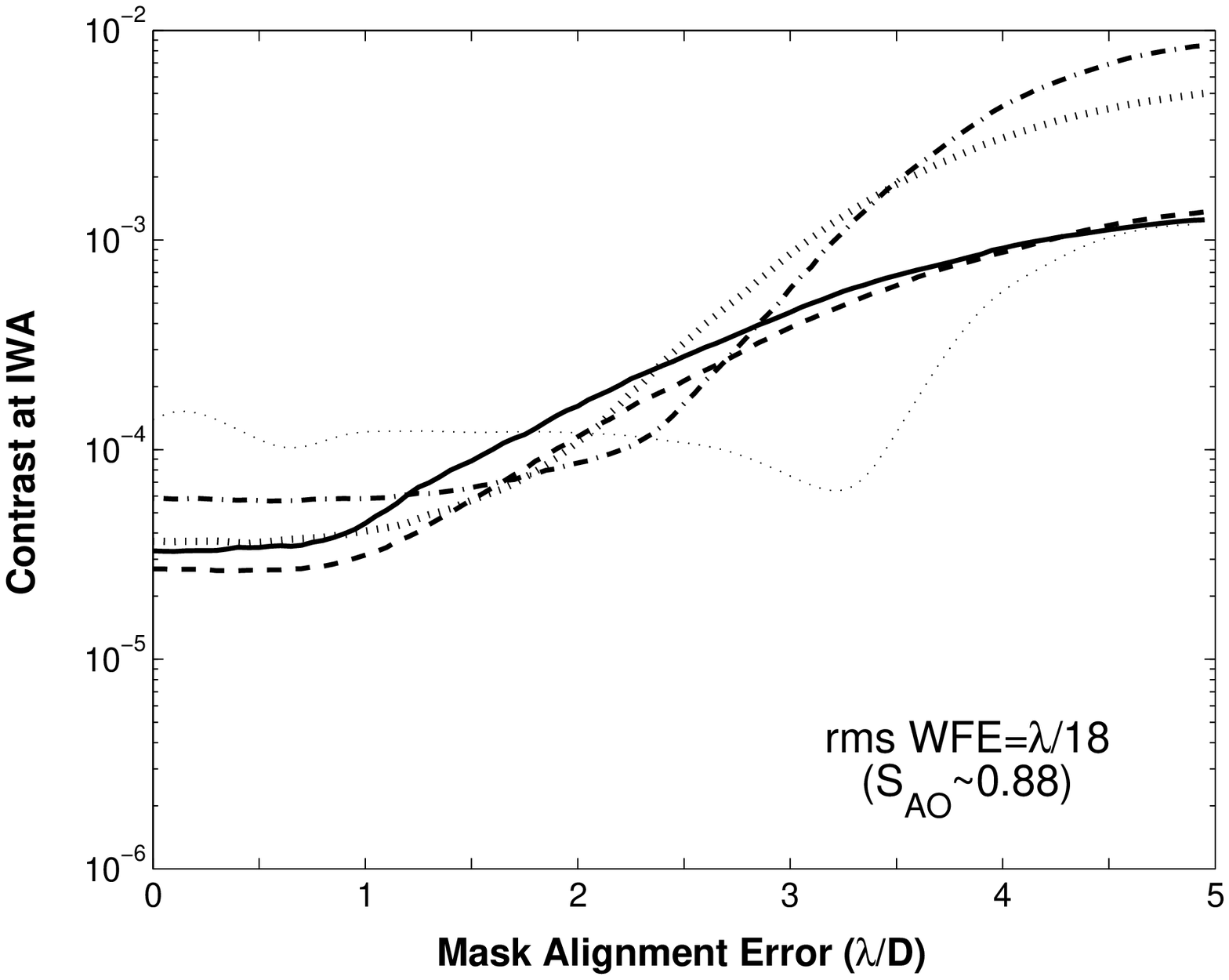}
\hfill
\includegraphics[width=3.1in]{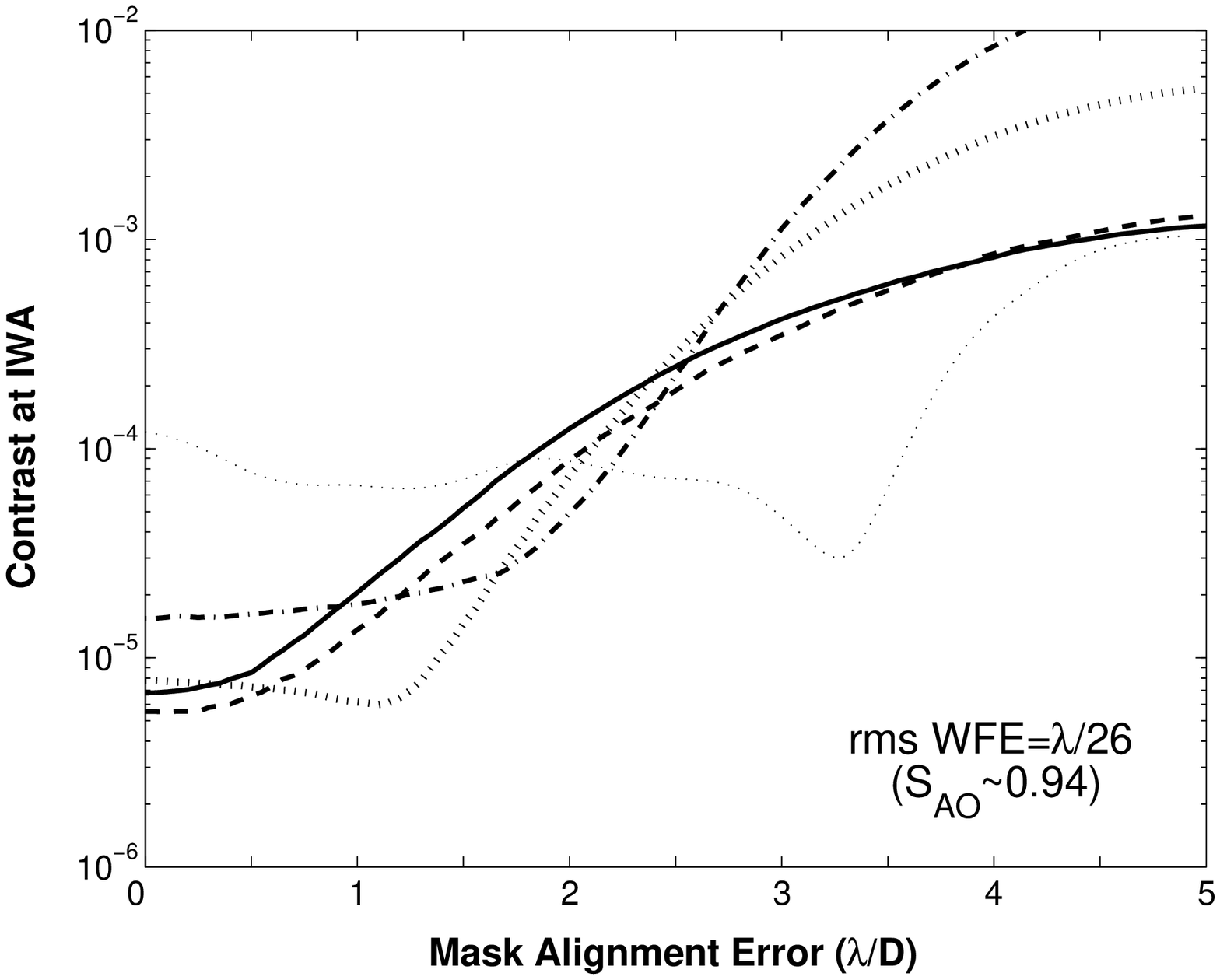}}
\caption{Median contrast within the $1 \, \lambda/D$ annulus outside
the coronagraph IWA as a function of systematic tilt error for
several characteristic levels of wavefront correction using optimal
Lyot stop sizes. Note that the Gaussian mask is a 4th-order mask.
The non-monotonic changes in contrast are a result of calculating
the median intensity with alignment errors in a circular geometry
and phasing between the Airy pattern and the mask intensity
transmission (see Lloyd \& Sivaramakrishnan 2005).} \label{fig:tt}
\end{figure}

This analysis is also applicable to low-order aberrations in a more
general sense. Shaklan \& Green 2005 have shown that the `order' of
the mask (4th, 8th, 12th, ... etc.) uniquely determines a
coronagraph's sensitivity to aberrations (tip/tilt, focus,
astigmatism, coma, trefoil, spherical, ... etc.). The result is that
higher-order masks, which are intrinsically broader, are naturally
better filters of any given low-spatial-frequency phase error. For
example, expansion of Equ.~\ref{equ:gaussian} shows that the
Gaussian mask intensity transmission profile near the optical axis
depends on $r$ raised to the fourth power; thus, it is a 4th-order
mask (Fig.~\ref{fig:masks}). Figure~\ref{fig:tt} confirms that the
Gaussian mask follows the 4th-order BL mask tilt sensitivity curve,
and that higher-order BL masks follow suit.

The hard-edge mask may be considered in the limit as the exponent of
the intensity transmission approaches infinity. It is effectively a
mask of infinite order, and thus the most resistant to
low-spatial-frequency aberration content. The sharp boundaries of
the hard-edge mask however also make the coronagraph leak the most
starlight. Combining this information, we recognize that
Fig.~\ref{fig:tt} is qualitatively illustrative of a trend
applicable to all individual phase aberrations, and sums of
(orthogonal) phase aberrations, introduced downstream from the AO DM
and wavefront sensor. As the phase errors increase, the contrast
generated by apodized masks will rise (degrade) from the AO noise
floor and eventually intersect the hard-edge mask curve.

In general, quasi-static phase errors further reduce the Strehl
ratio. Therefore the situation is slightly more complicated than
with tip/tilt alone, which, strictly speaking, is a change to the
pointing vector and not an aberration. Nevertheless, we can extend
the principle to quantitatively include them all.



Consider the final measured Strehl ratio, $S$, written as a
decomposition of uncorrelated errors (Sandler et al. 1994):
\begin{equation}
S=S_1 \: S_2 \: S_3 \; ... \; S_n,
\end{equation}
where each $S_{i}$ with $1 \leq i \leq n$ represents an independent
Strehl degradation. An equivalent statement is that uncorrelated
wavefront errors add in quadrature. Since the Strehl ratio produced
by the AO system is unrelated to the subsequent optical path, we let
$S=S_{AO} \: S_{qs}$ describe the final stellar image, where
$S_{AO}$ is the AO-corrected Strehl ratio (as has been used
throughout the text and figures) and $S_{qs}$ is the Strehl ratio
due to {\it all} quasi-static phase aberrations introduced
downstream from the AO DM and wavefront sensor.

Results from $\S\ref{sec:hard_vs_apo}$ and Fig.~\ref{fig:tt} show
that use of an apodized mask, preferably the 4th-order BL mask
($\S\ref{sec:gauss_vs_bl}$), is justified only when $S_{AO} \gtrsim
88\%$. Thus, we require that the measured Strehl ratio satisfy the
condition:
\begin{equation}
S \gtrsim 0.88 \: S_{qs},
\label{equ:criterion}
\end{equation}
where $0 \leq S_{qs} \leq 1$ is the intrinsic Strehl of the optical
system. With an internal fiber-coupled source, such as a calibration
lamp, $S_{qs}$ can be measured when the AO DM is inactive and flat,
so far as one might trust zeroing the actuator voltages. We note
that this relationship is valid only in the high-Strehl regime
($\sigma \lesssim \lambda / 2 \pi$).


The effects of speckle-nulling can be incorporated by noticing that
contrast curves such as those shown in Fig.~\ref{fig:tt}, and
similar graphs for the other phase errors present in a real system,
will be ``flattened" by the additional DMs (see
$\S\ref{sec:qs_aberr}$). In other words, the extra degrees of
freedom afforded with this hardware compensate for quasi-static
aberrations whose spatial-frequencies match the intended search
area. Long-lived speckles may be nulled to an intensity level where
the contrast curves in Fig.~\ref{fig:tt} intersect the vertical
axes; this is true for any such aberration. A candidate companion
would then be noticed by the inability of the instrument to remove
its mutually incoherent signal. Subsequent changes to the shape of
the DMs and hence location of the dark hole might indicate the
presence of other faint sources.


As an example, consider a measured Strehl ratio $S=85\%$ with a
system that has an intrinsic Strehl ratio $S_{qs}=94\%$. According
to Equ.~\ref{equ:criterion}, we are indeed justified in using the
4th-order BL mask for this application. However, if the additional
DMs cannot remove quasi-static speckles from the region of interest
down to the intensity of the AO-limited noise floor, we may consider
switching to a higher-order mask, such as the 8th-order or
12th-order BL mask, to help filter stellar residuals before they
illuminate the detector.






These results allow us to make rather strong conclusions regarding
the implementation guidelines for image masks included in this
study. We state them concisely in the next section. For further
discussion of low-order phase aberrations within the context of Lyot
coronagraphy, see Sivaramakrishnan et al. 2005, Lloyd \&
Sivaramakrishnan 2005, Shaklan \& Green 2005, and Crepp et al. 2006.





\section{Summary \& Concluding Remarks}
\label{sec:conclusions}


One should select an image mask whose rejection of starlight is
commensurate with the noise floor set by the AO system. Our numerical
simulations imply the following: \\

{\it (1) Apodized masks should replace the hard-edge mask only when
on-sky Strehl ratios, $S$, exceed $\sim 0.88 \: S_{qs}$, where
$S_{qs}$ is the Strehl ratio due to {\it all} quasi-static phase
aberrations introduced by the instrument, downstream from the AO DM.
Below this level of correction, the hard-edge mask outperforms
apodized masks, not by reaching deeper levels of contrast, but by
generating similar contrast ($\S\ref{sec:hard_vs_apo}$) and
throughput ($\S\ref{sec:gauss_vs_bl}$) with more resistance to
quasi-static errors. This result is independent of entrance aperture
geometry and bandpass.






(2) Since 4th-order BL masks yield more Lyot stop throughput than
the Gaussian mask, and apodized masks with smooth intensity
transmission gradients are equally difficult to manufacture,
Gaussian masks should not be implemented under any foreseeable
conditions on telescopes with uniform transmission entrance
apertures.


(3) The selection of higher-order BL masks over the 4th-order BL
mask is relegated to situations where both the ability to correct
for the atmosphere to a very high degree and the {\it in}ability to
adequately null quasi-static speckles is simultaneously
present.} \\

For the operating range often considered with a traditional Lyot
coronagraph, IWA $= 3-5 \,\, \lambda_{max} / D$, the exact contrast
and throughput values will deviate from those reported in
$\S$\ref{sec:comparative}, but in a rather predictable manner. The
italicized conclusions however do not change with these
considerations. It is also important to mention that Strehl ratios
can be somewhat pesky to calculate in practice. Experimentally
determined values are accurate only to several percent if the image
is not spatially sampled at a rate higher than twice the Nyquist
frequency (Roberts et al. 2004).






The hard-edge occulting mask is a remarkably relevant coronagraphic
tool in an age of sophisticated wavefront correction techniques and
clever applications of Fourier optics. Apodized masks, or binary
versions of apodized masks (Kuchner \& Spergel 2003), require
nano-fabrication capabilities at visible and near-IR wavelengths
(Balasubramanian et al. 2006, Crepp et al. 2006, Carson et al.
2005b, Debes et al. 2004, Trauger et al. 2004); this can be a strong
deterrent and should be avoided unless Equ.~\ref{equ:criterion} is
satisfied (although, in a fast optical system, the focal ratio may
impose stringent tolerances when building a hard-edge mask as well).
In the realm of extreme AO, the 4th-order BL mask should be
implemented, so long as quasi-static phase aberrations are
manageable with speckle-nulling hardware.


Finally, our calculations suggest that the wavefront correction
levels required for ground-based observations preclude reliable
spectroscopic measurements of close-separation companions that are
more than approximately one million times dimmer than their host
star. Presumably, space-based instruments will be able to do much
better. Nevertheless, this result is still more than an order of
magnitude deeper than current AO-coronagraphs provide. \\




The authors are grateful to Joseph Carson for his guidance in AO
simulation code development and the anonymous referee for helpful
comments that improved the paper. This work was partially supported
by NASA with grants NNG06GC49G, NNG05G321G and NNG05GR41G, the NSF
with grants NSF AST04-51407 and AST04-51408, the TPF program, the
UCF-UF SRI program, and the University of Florida. JC acknowledges
support from the SPIE and a Grant-In-Aid-of-Research from the
National Academy of Sciences, Administered by Sigma Xi, The
Scientific Research Society.



\small{

}

\clearpage

\begin{table}[!hb]
\centerline{
\begin{tabular}{ccccc}
\hline
 & {\bf Mask}     &      {\bf Star$\times$Mask}     &      {\bf Lyot Plane}     & {\bf Final Image} \\
\hline
\raisebox{0.3in}[0pt]{{\bf Hard-Edge}} & \includegraphics[width=1in]{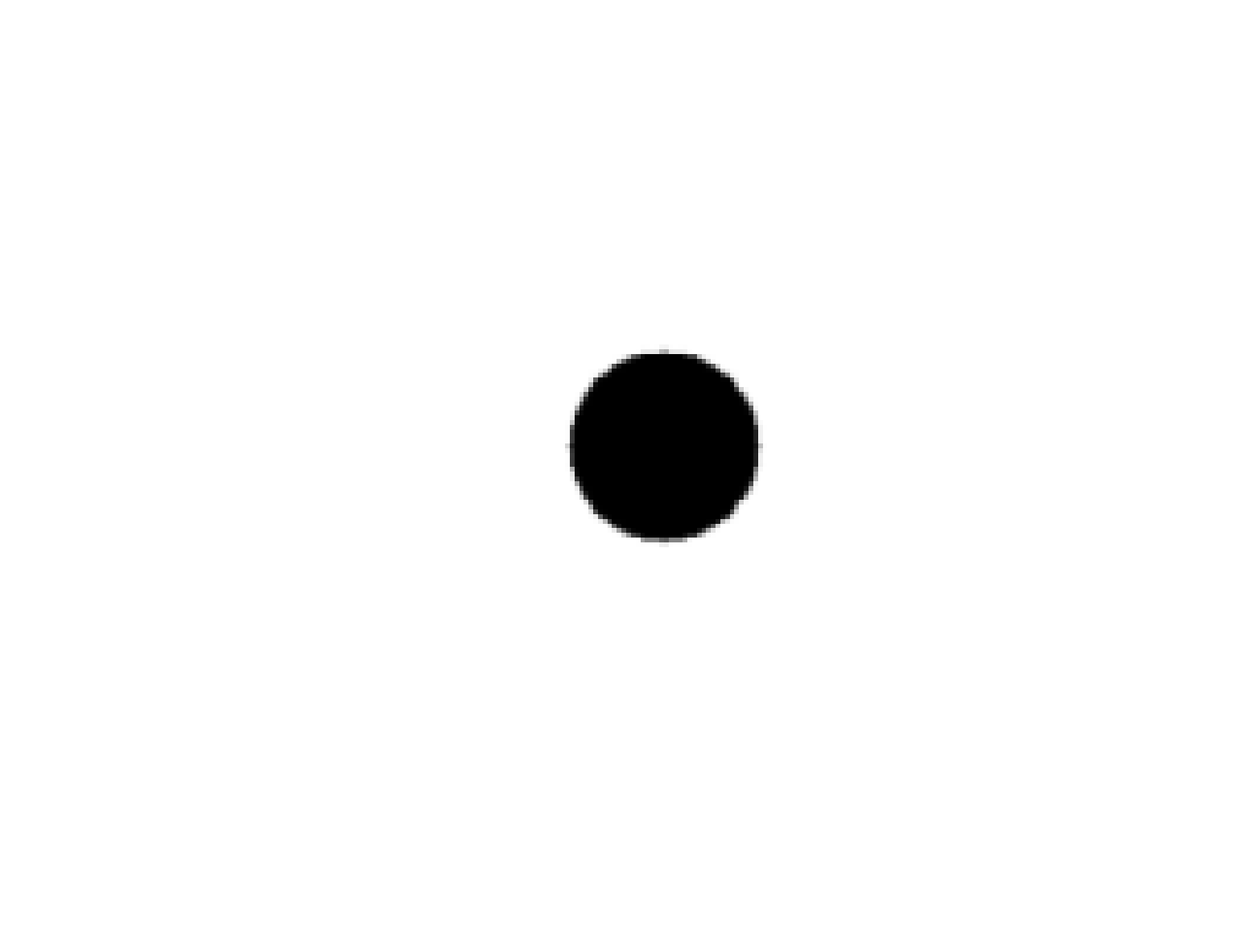} & \raisebox{-0.03in}[0pt]{\includegraphics[width=1in]{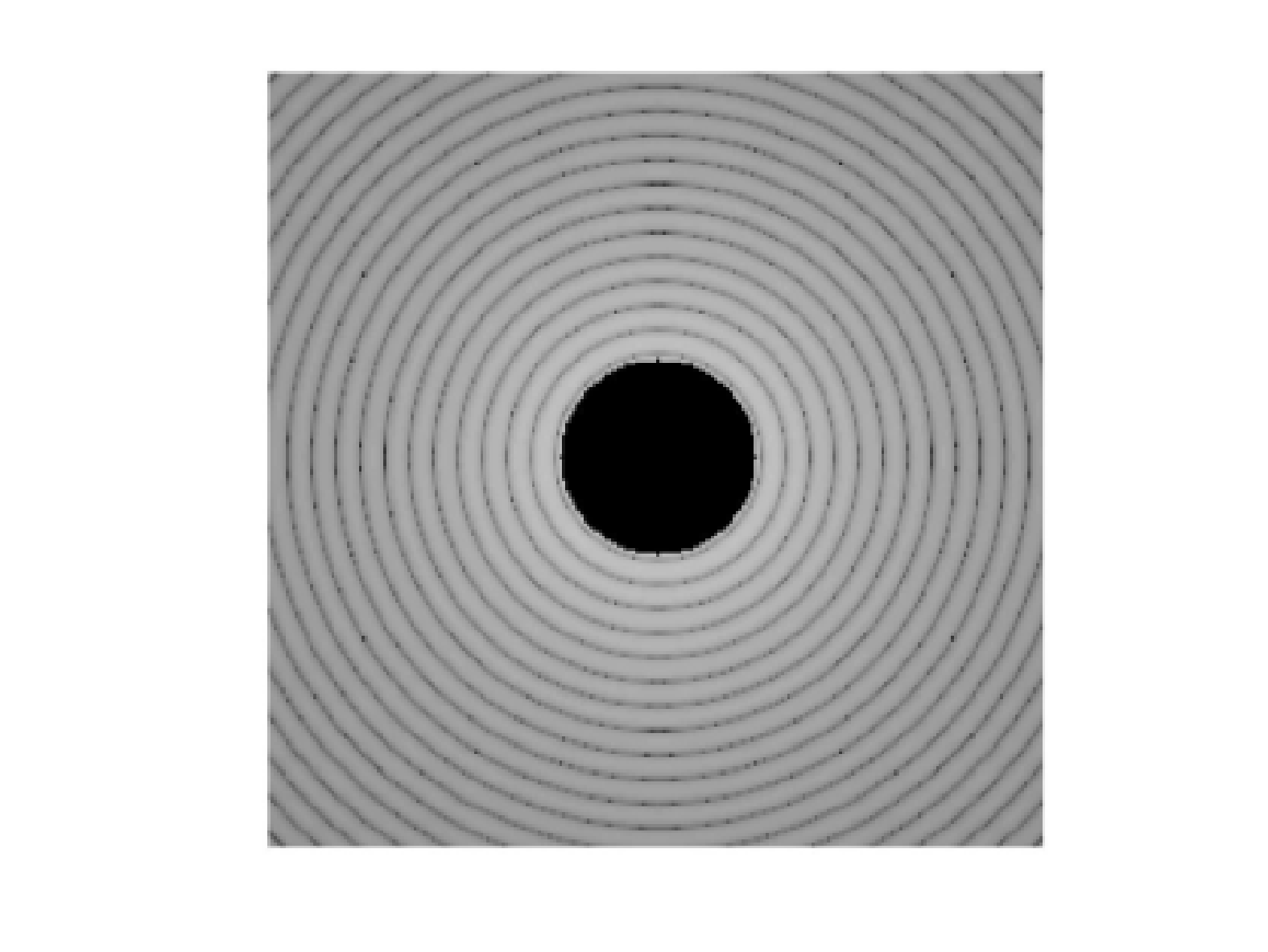}} & \raisebox{+0.041in}[0pt]{\includegraphics[width=0.632in]{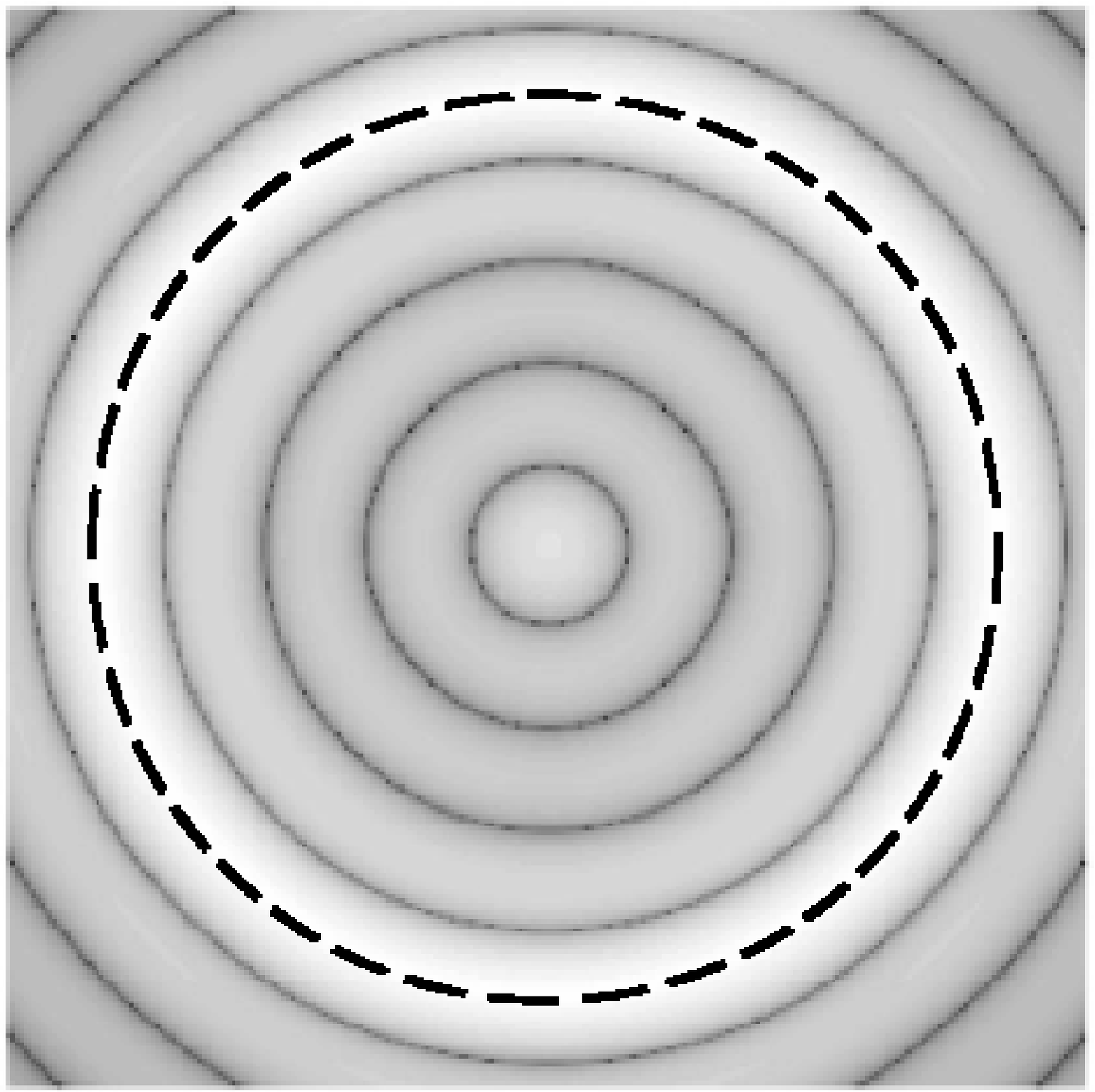}} & \hspace{0.665cm}\raisebox{+0.02in}[0pt]{\includegraphics[width=0.854in]{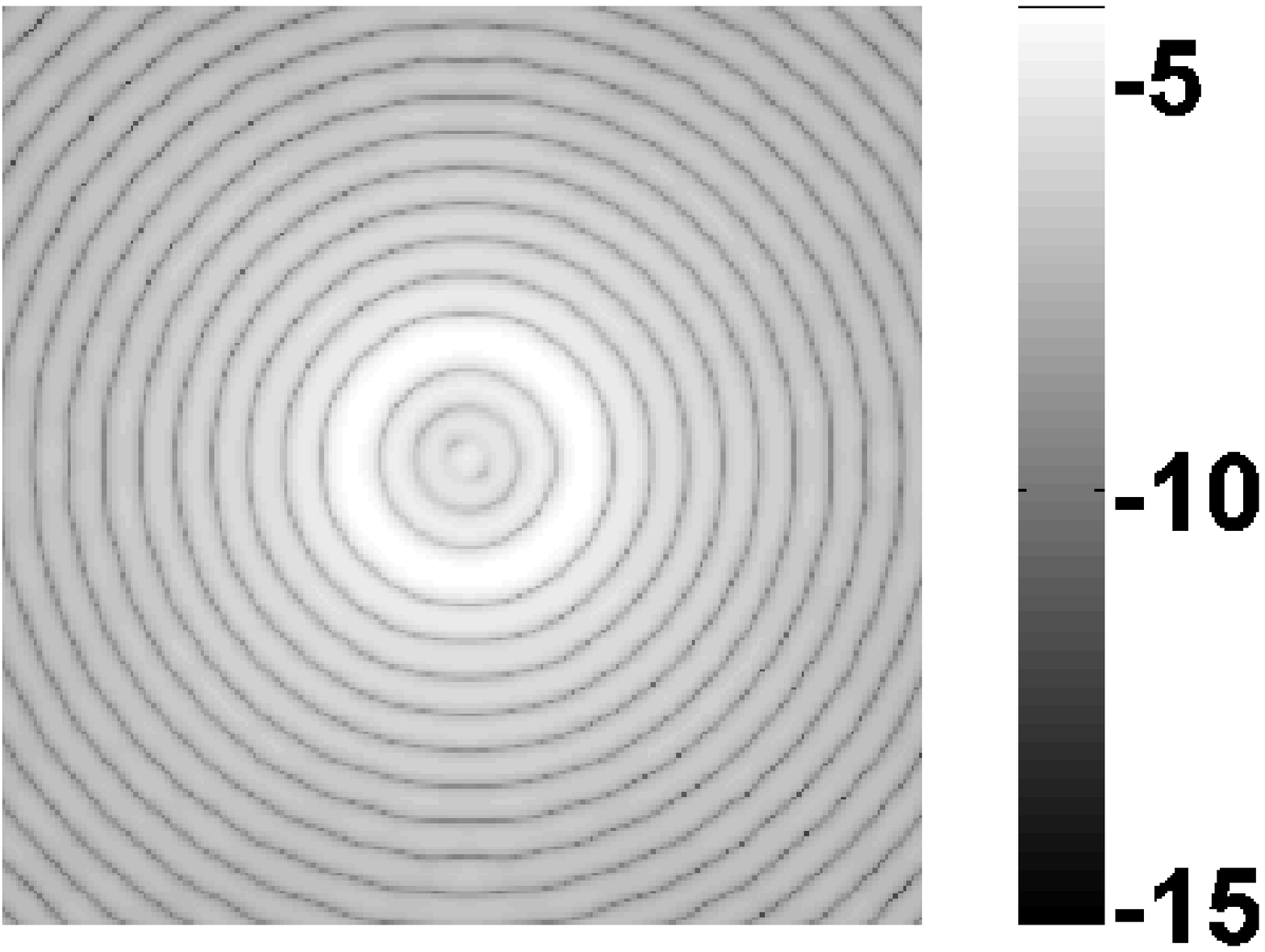}}\\
\hline
\raisebox{0.3in}[0pt]{{\bf Gaussian}} & \includegraphics[width=1in]{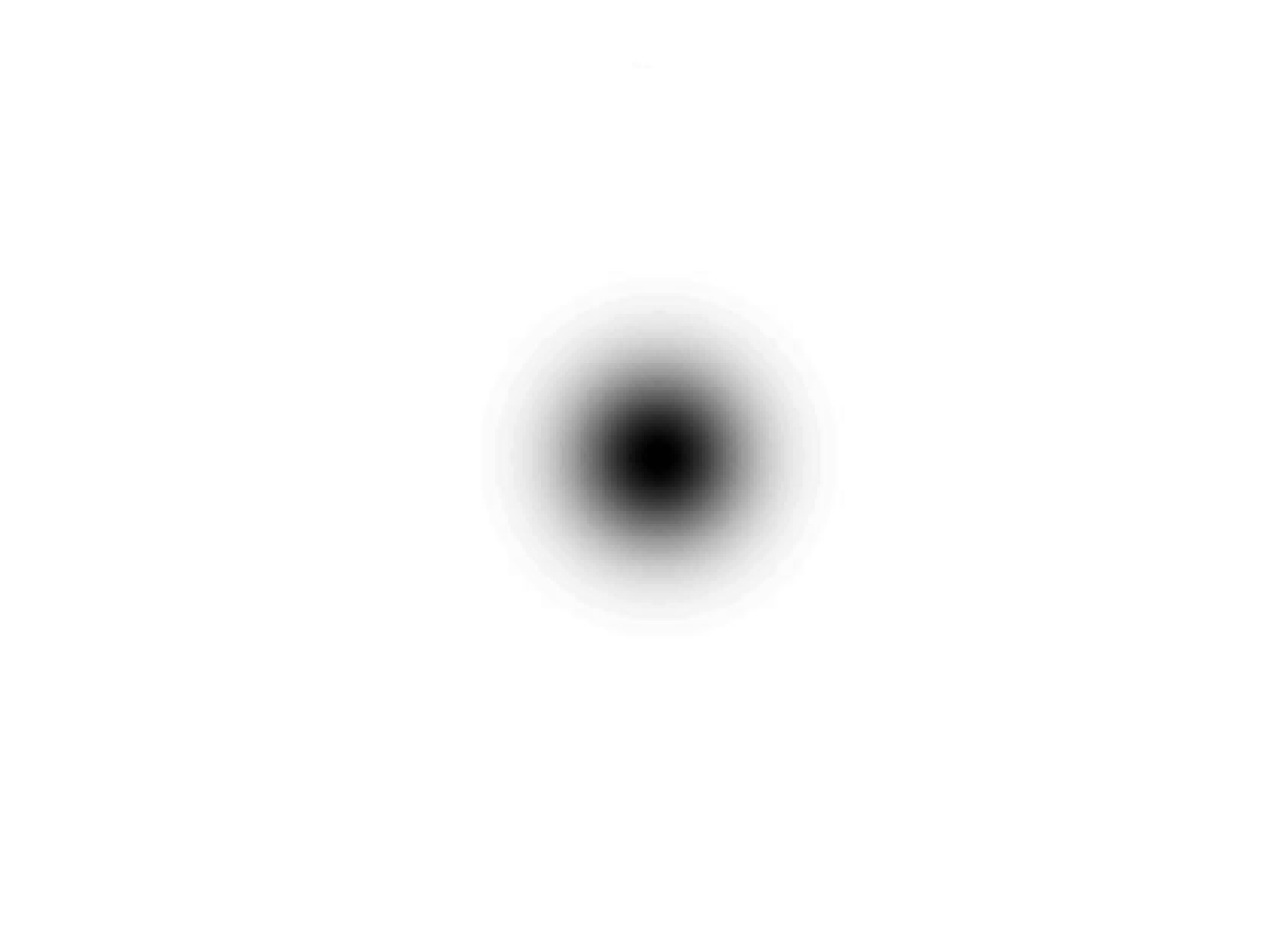} & \raisebox{-0.03in}[0pt]{\includegraphics[width=1in]{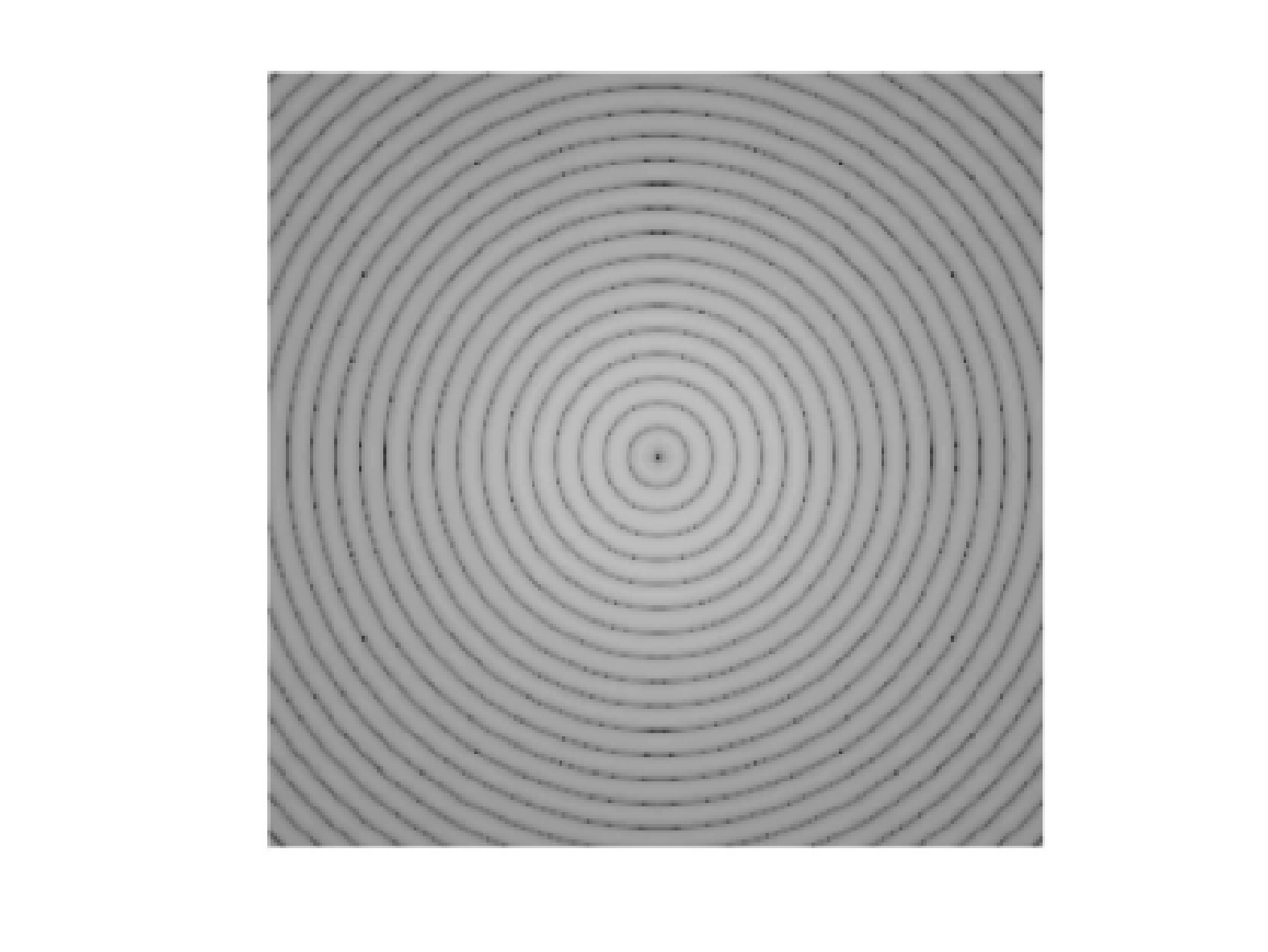}} & \raisebox{+0.041in}[0pt]{\includegraphics[width=0.632in]{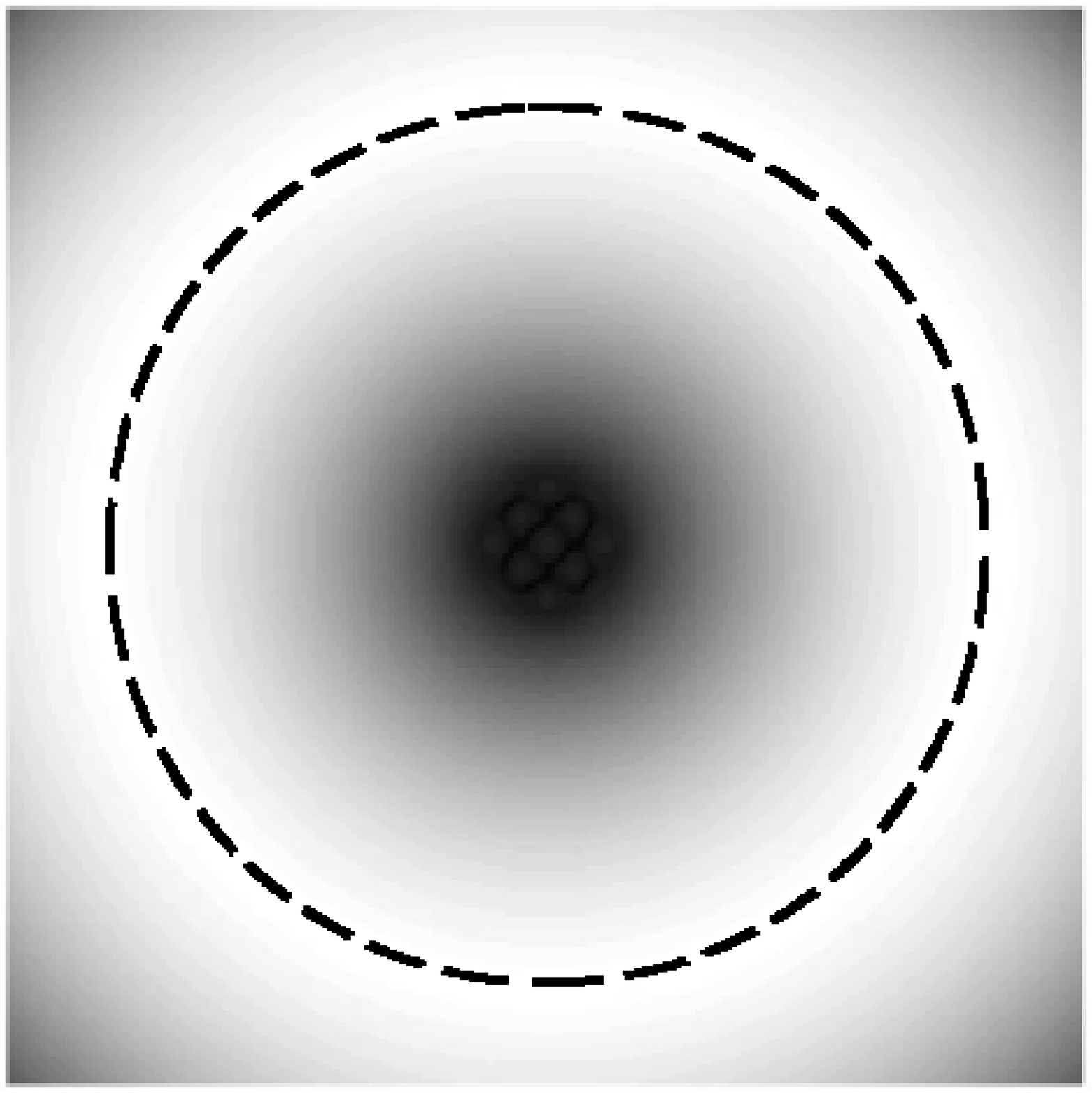}} & \raisebox{-0.03in}[0pt]{\includegraphics[width=1in]{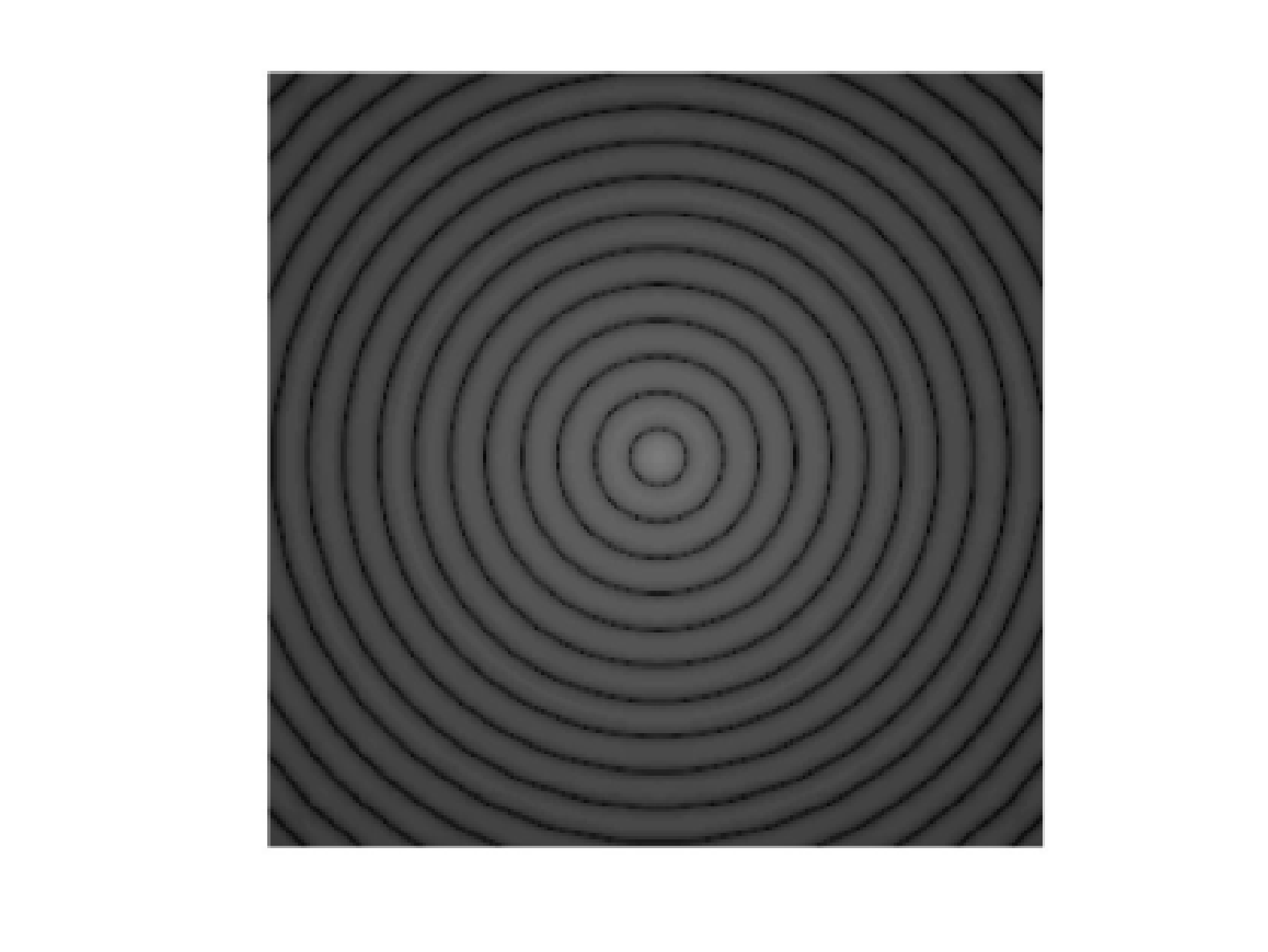}}\\
\hline
\raisebox{0.3in}[0pt]{{\bf 4th-order}} & \includegraphics[width=1in]{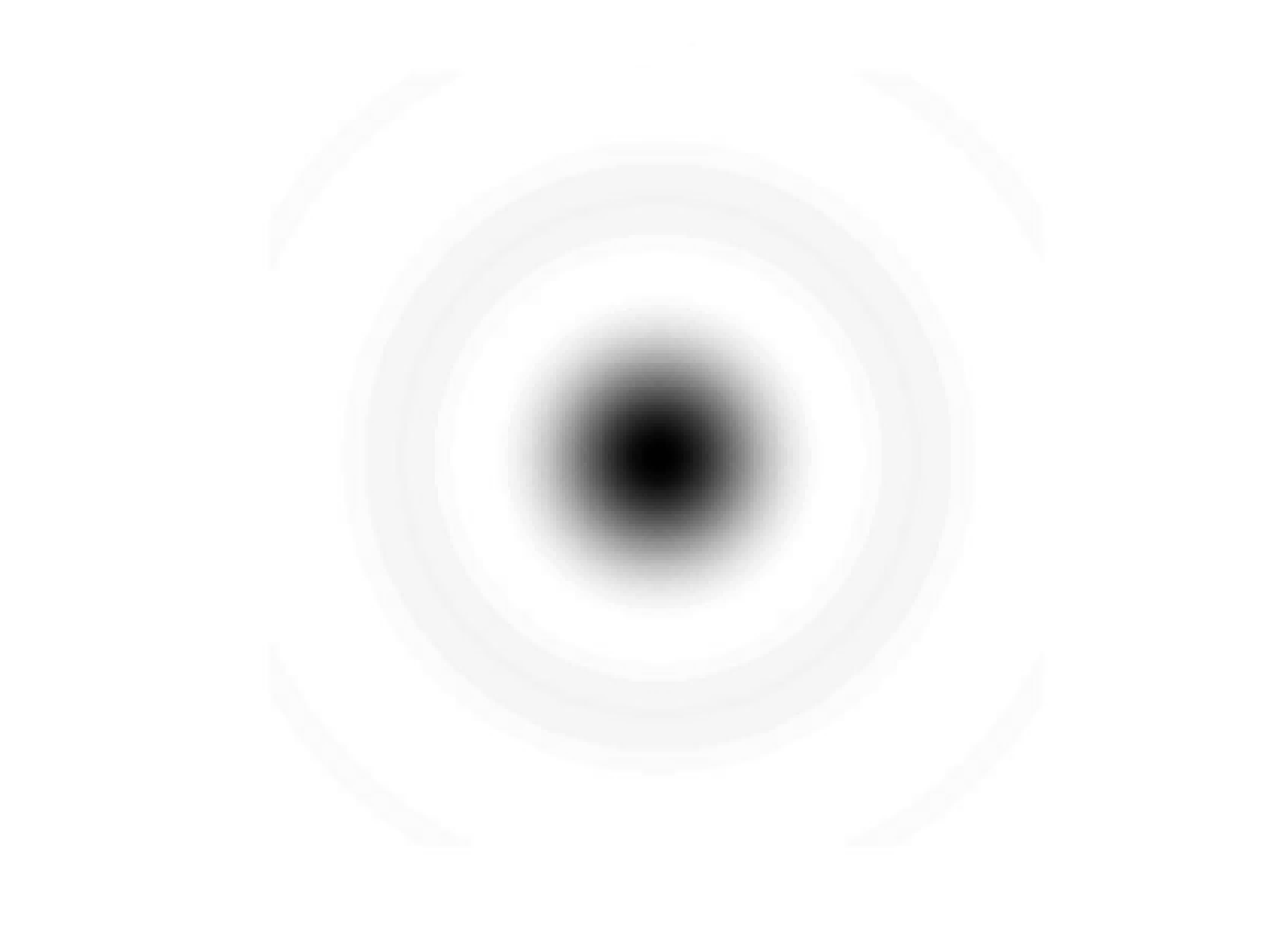} & \raisebox{-0.03in}[0pt]{\includegraphics[width=1in]{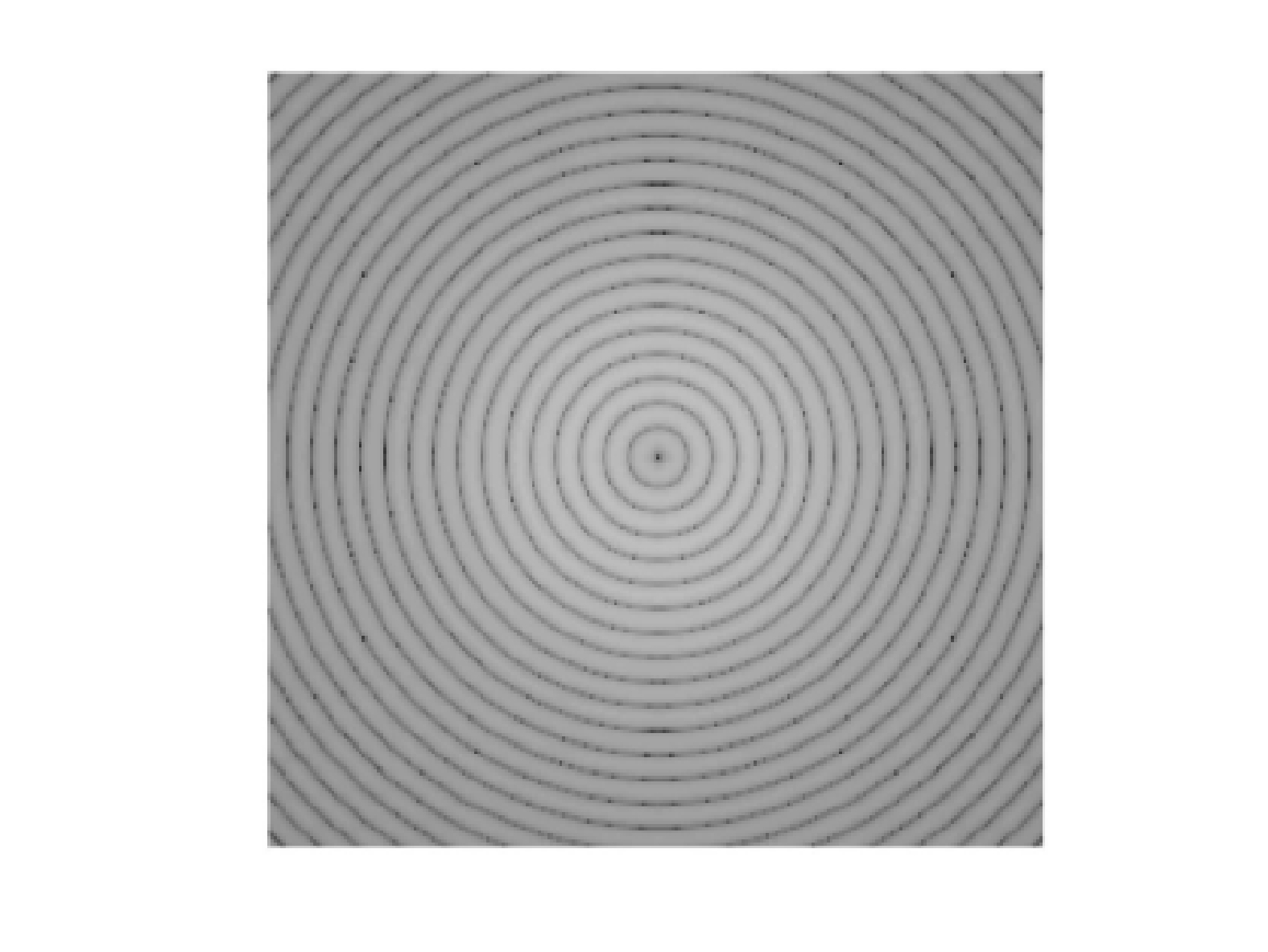}} & \raisebox{+0.041in}[0pt]{\includegraphics[width=0.632in]{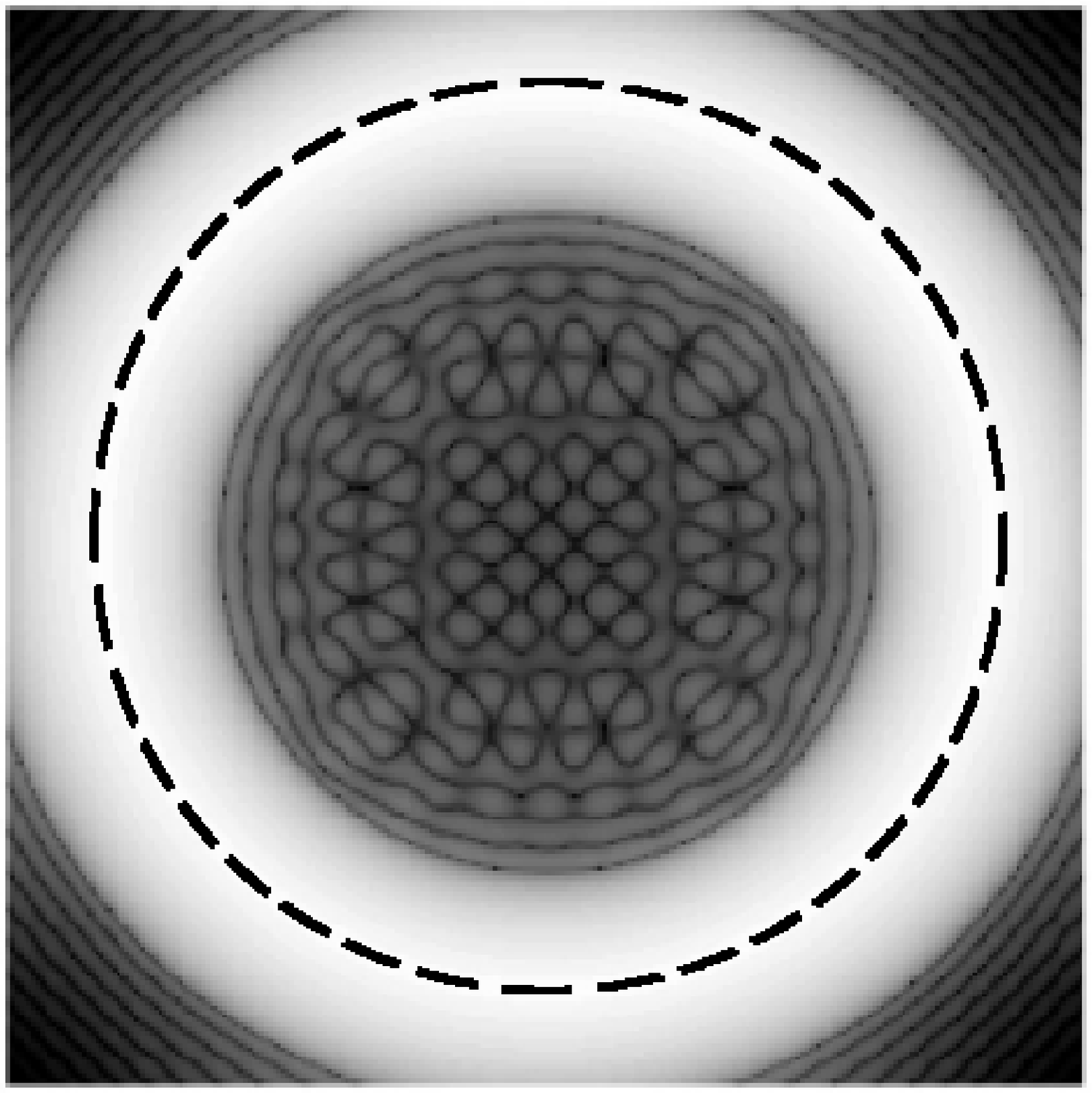}} & \raisebox{-0.03in}[0pt]{\includegraphics[width=1in]{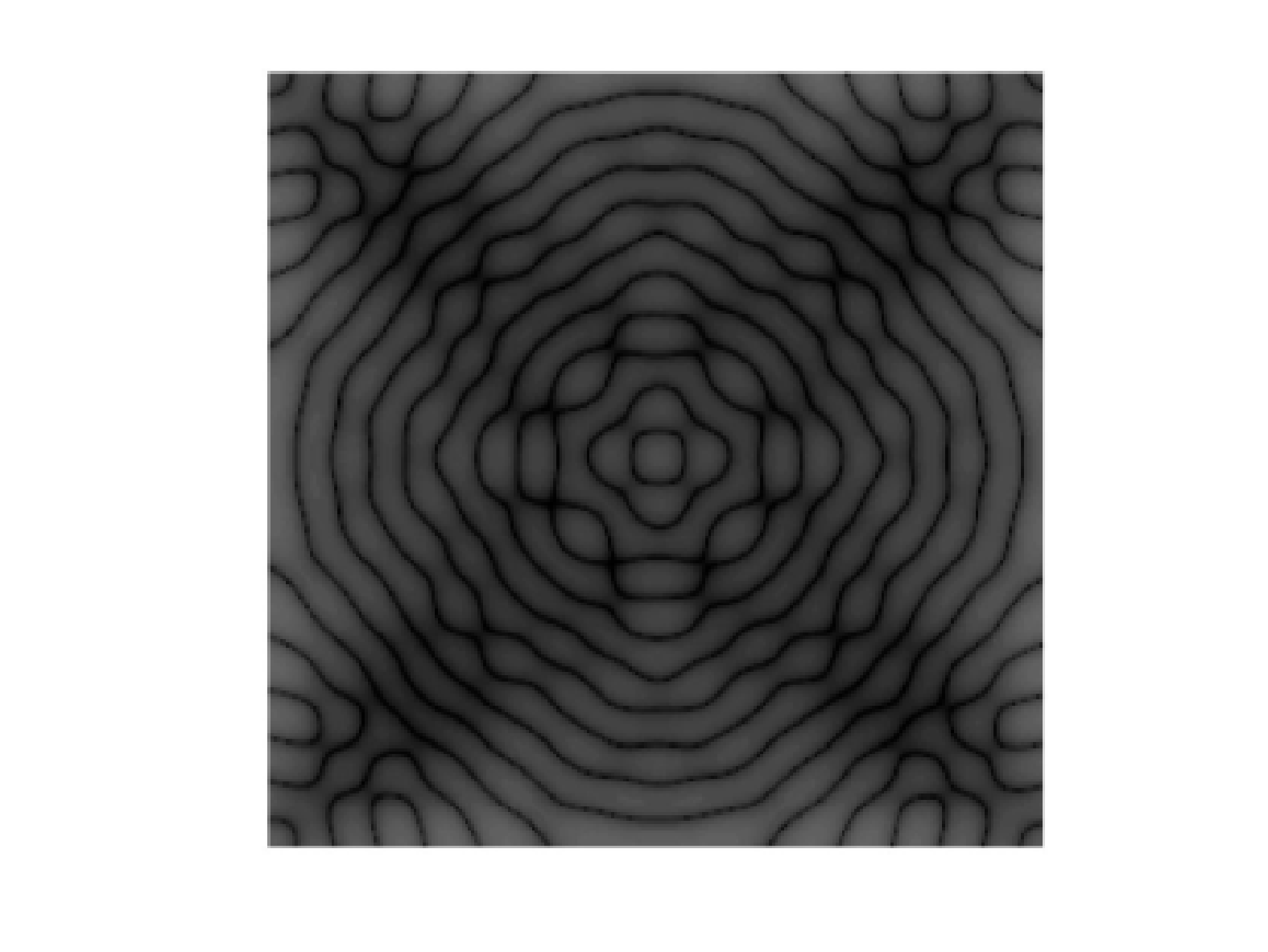}}\\
\hline
\raisebox{0.3in}[0pt]{{\bf 8th-order}} & \includegraphics[width=1in]{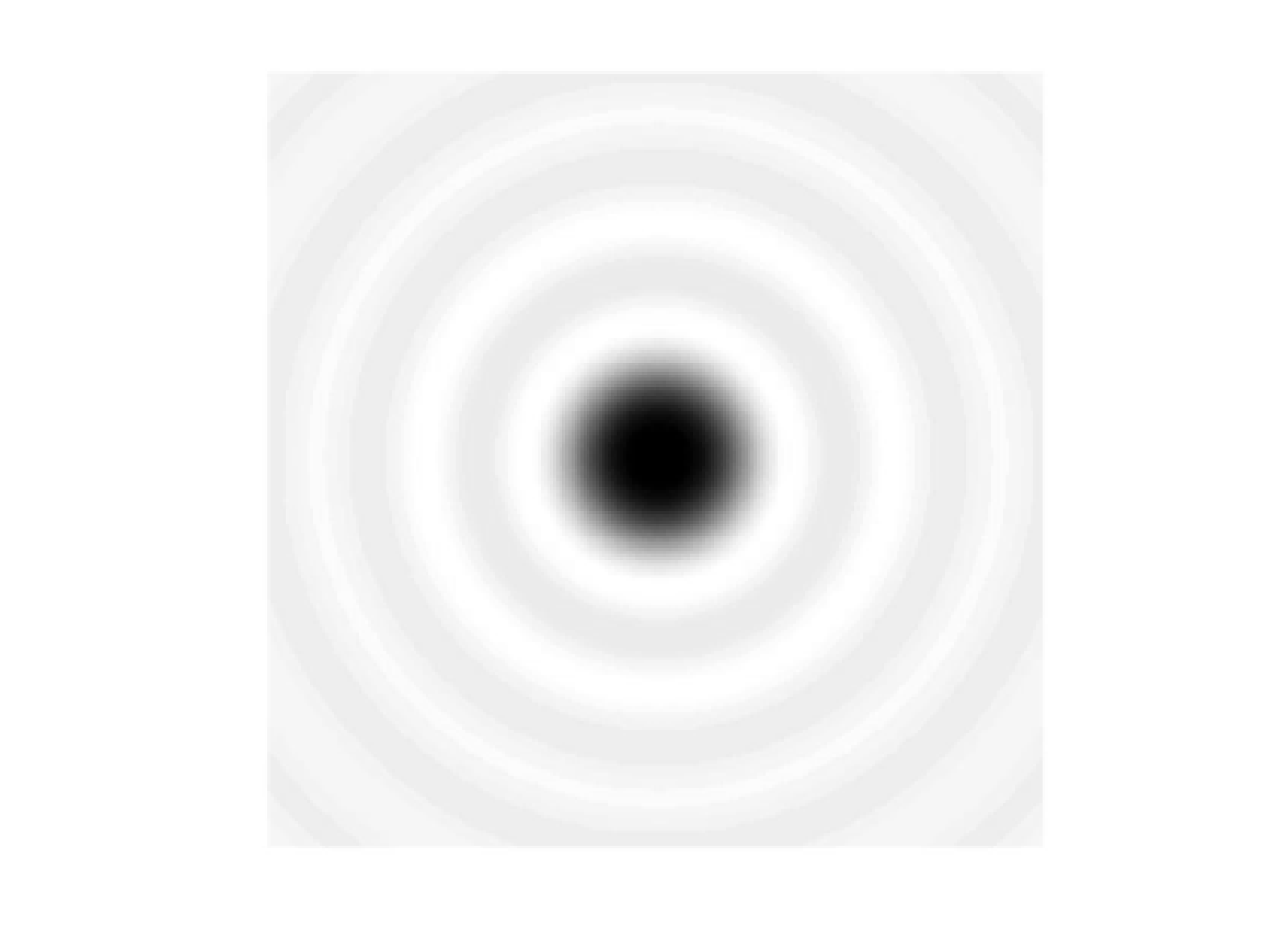} & \raisebox{-0.03in}[0pt]{\includegraphics[width=1in]{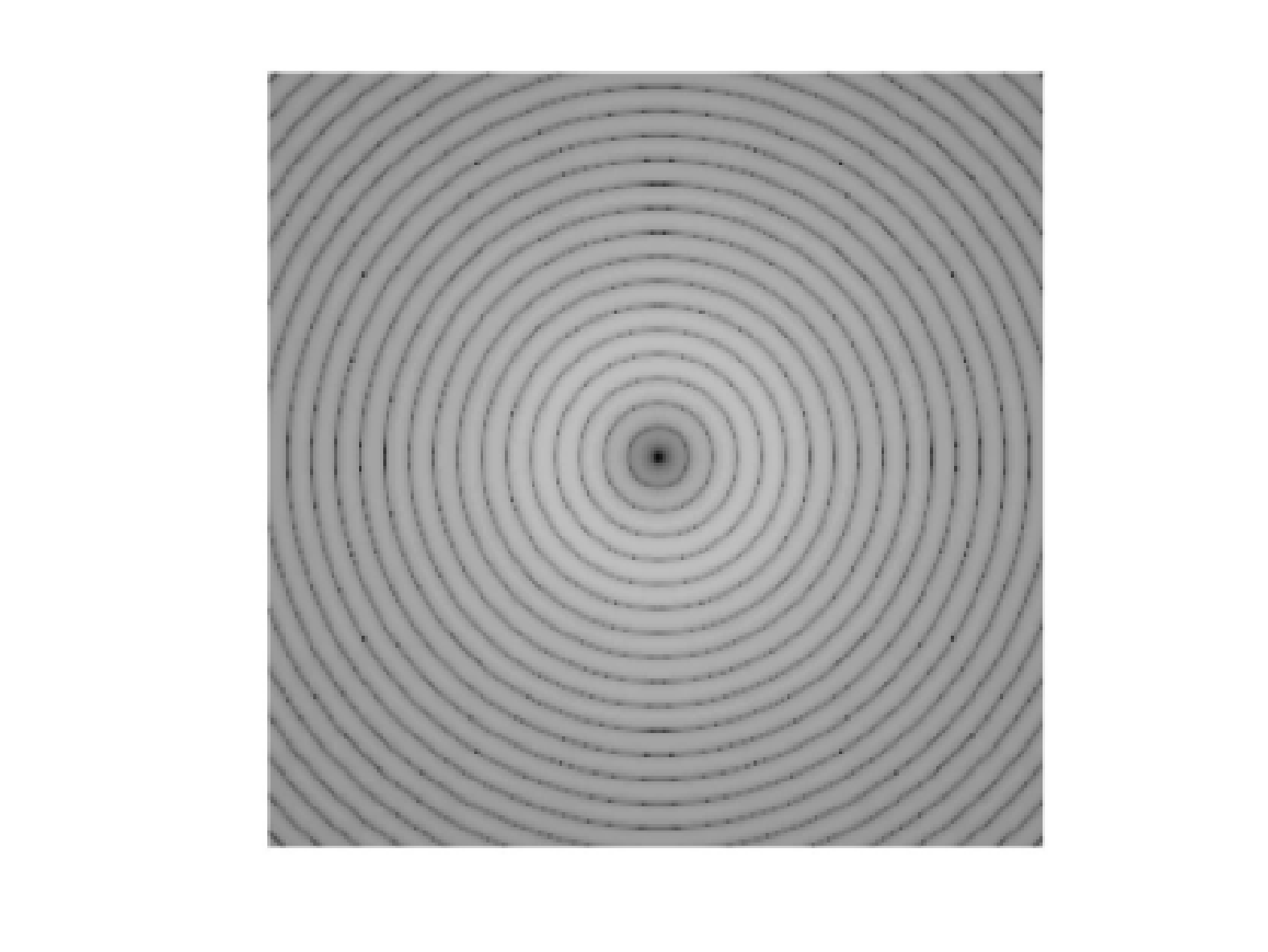}} & \raisebox{+0.041in}[0pt]{\includegraphics[width=0.632in]{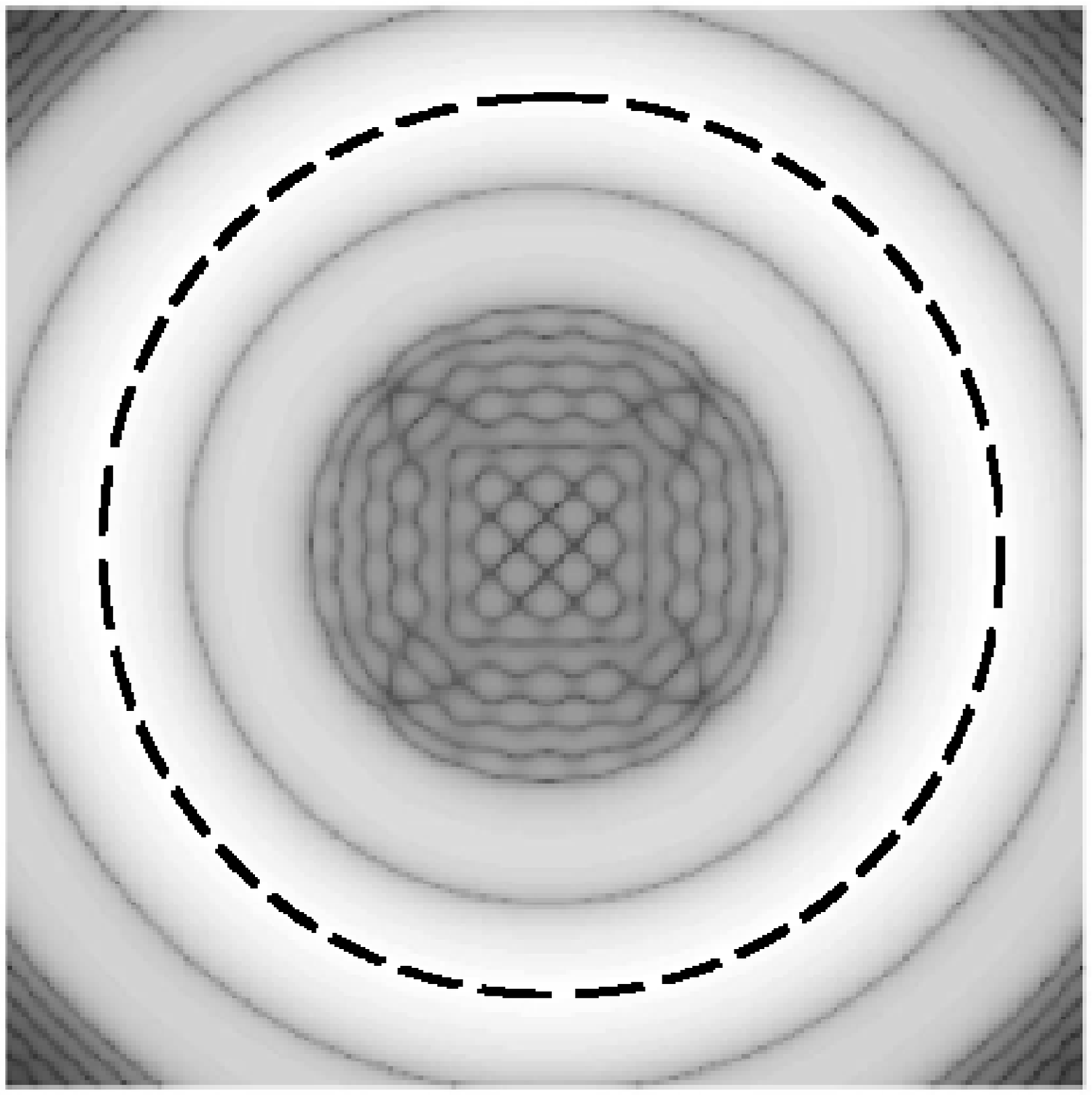}} & \raisebox{-0.03in}[0pt]{\includegraphics[width=1in]{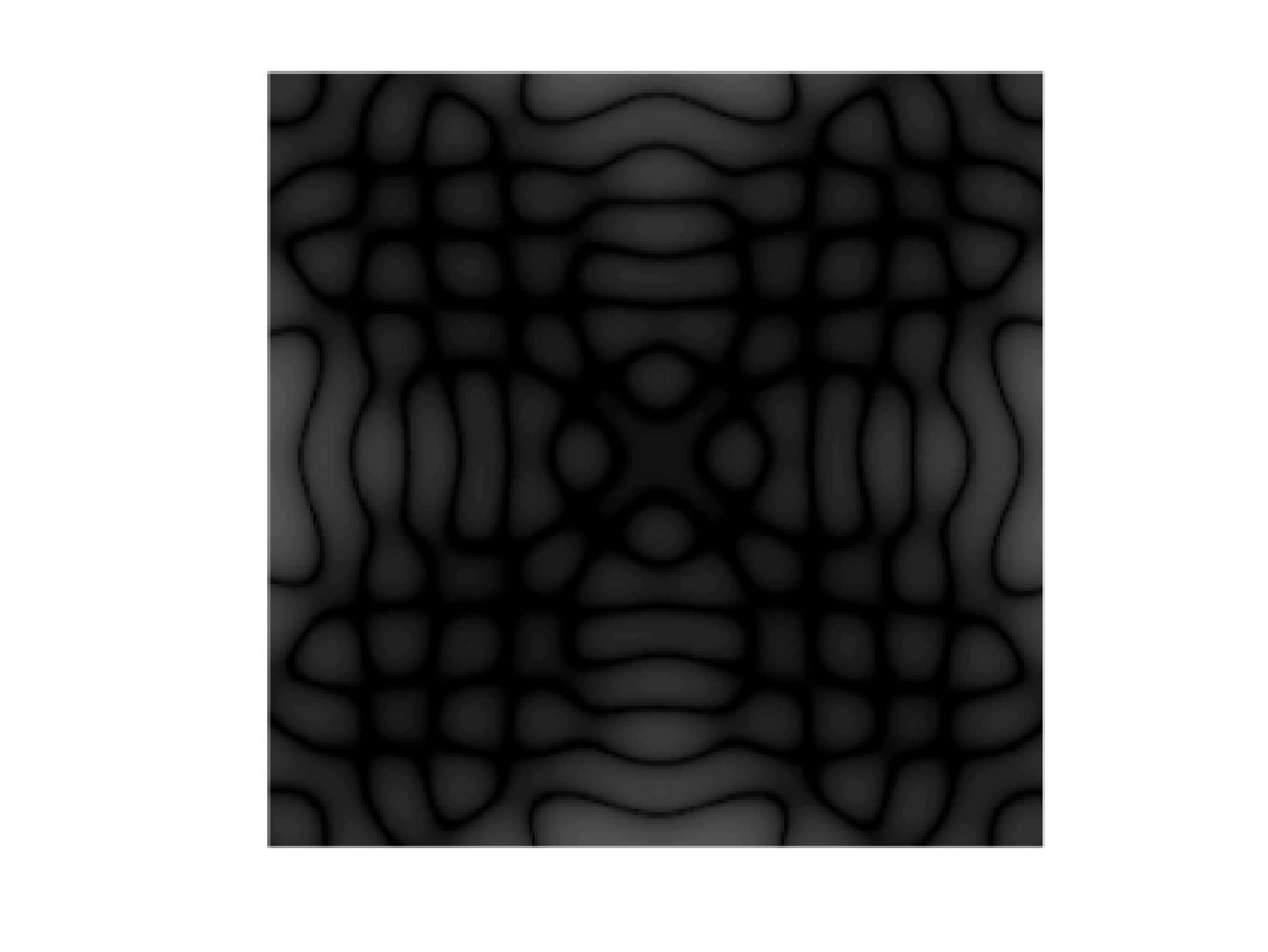}}\\
\hline
\raisebox{0.3in}[0pt]{{\bf 12th-order}} & \includegraphics[width=1in]{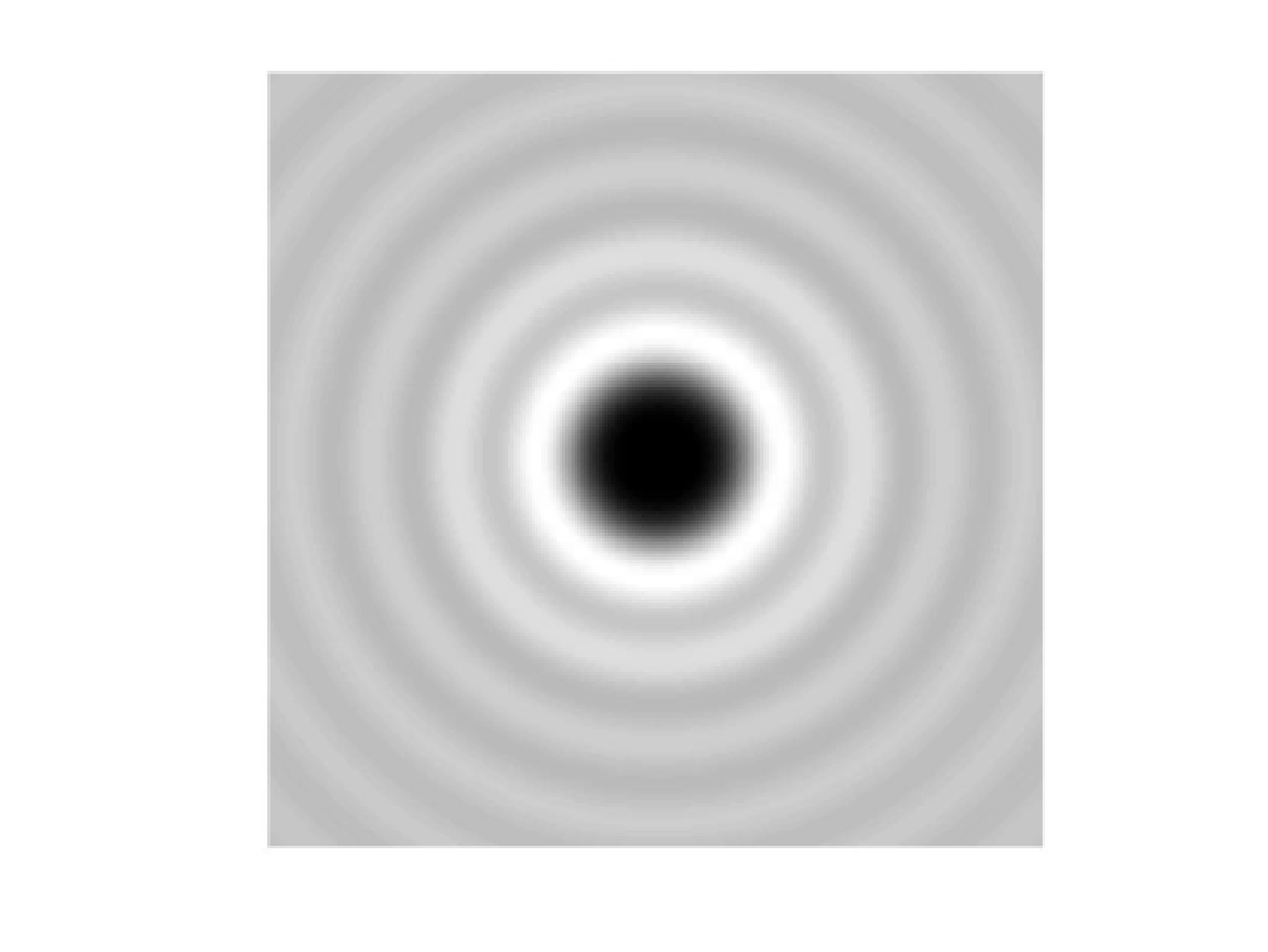} & \raisebox{-0.03in}[0pt]{\includegraphics[width=1in]{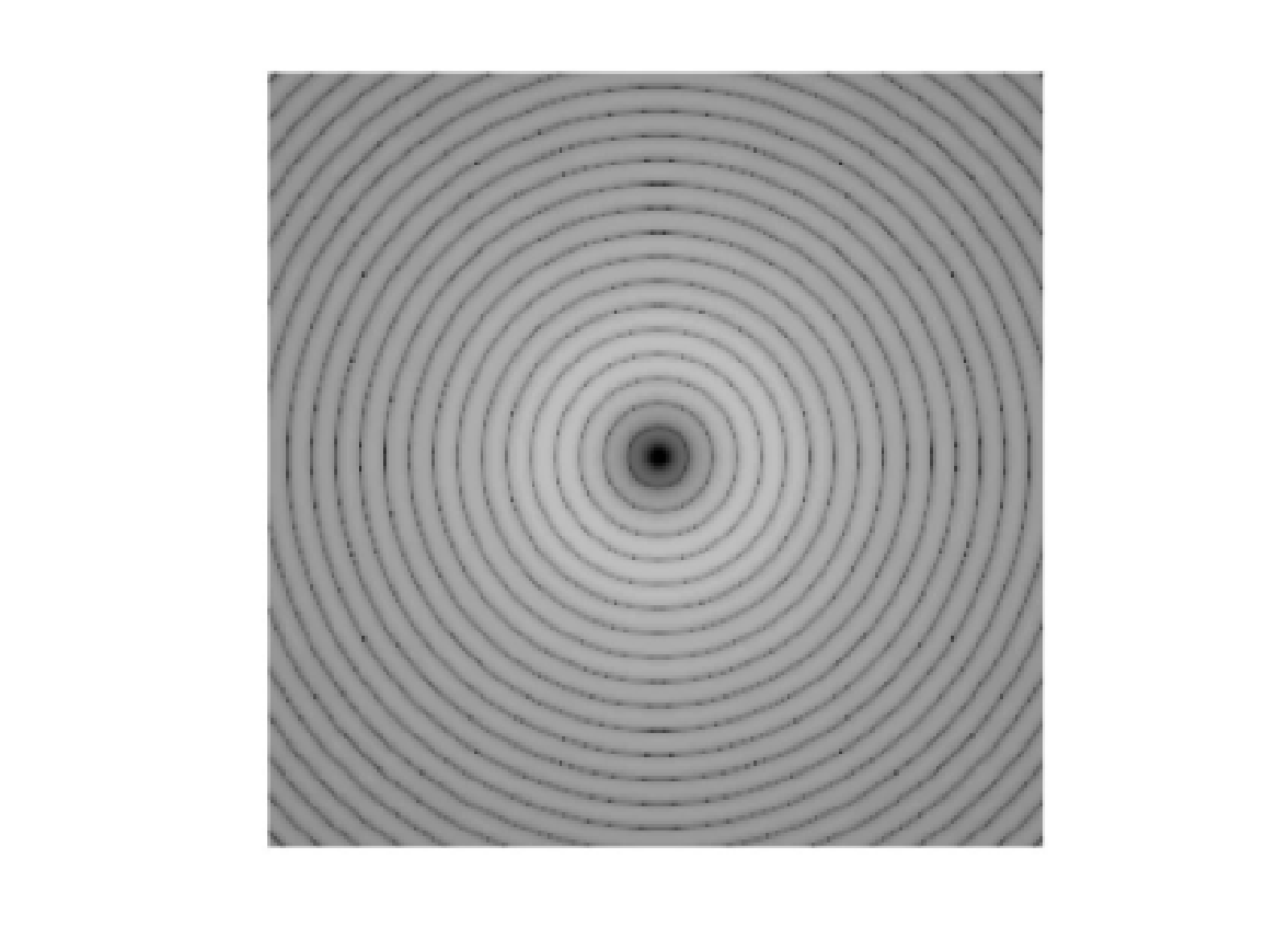}} & \raisebox{+0.041in}[0pt]{\includegraphics[width=0.632in]{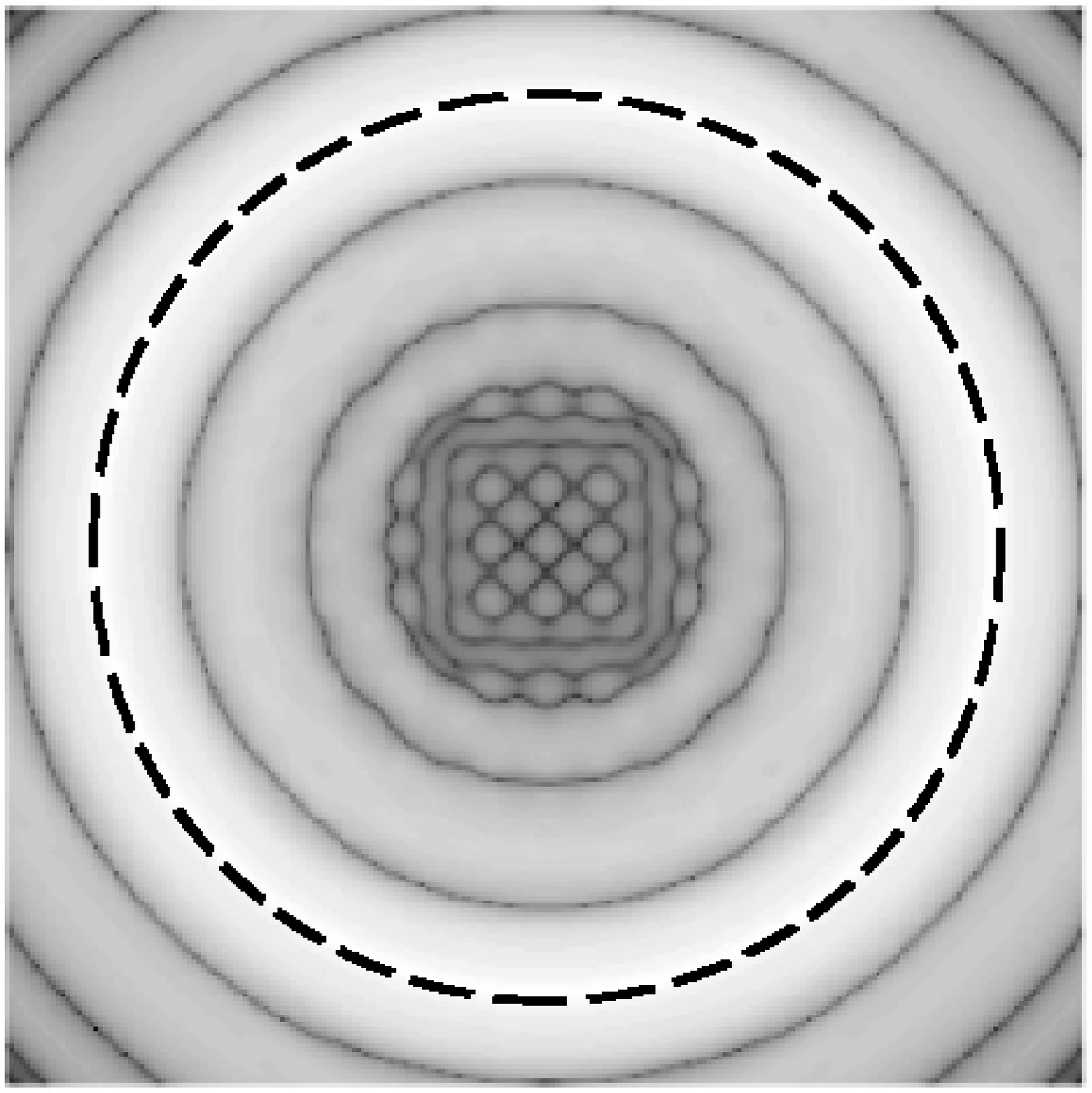}} & \raisebox{-0.03in}[0pt]{\includegraphics[width=1in]{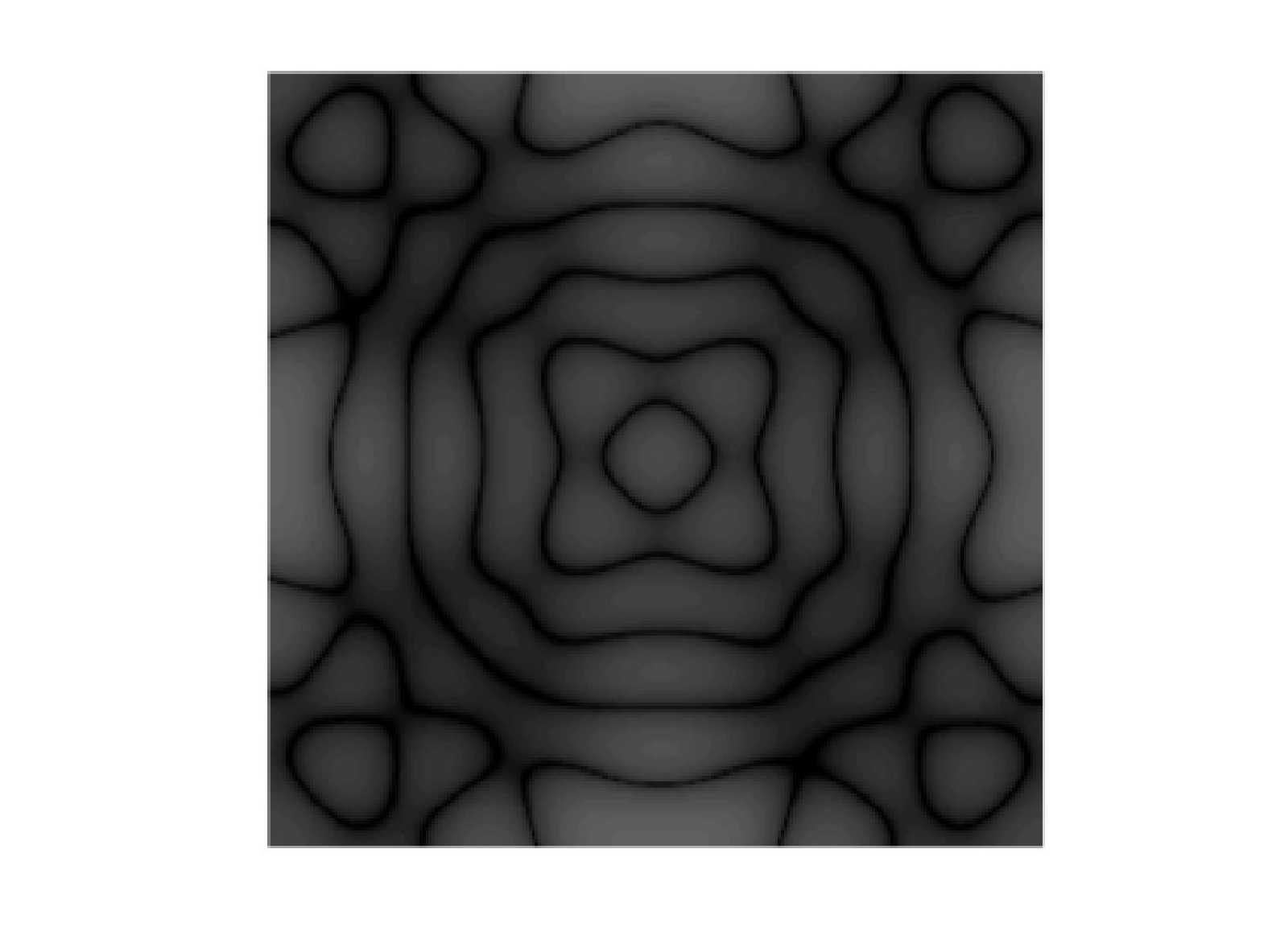}}\\
\hline
\end{tabular}}
\caption{Coronagraph simulations with perfect incident wavefronts.
Intensities in the first two columns are shown on the same
logarithmic scale using the mask profiles, $0 \leq |M(r)|^2 \leq 1$,
in Fig. 1 and a normalized Airy pattern. The spatial extent of the
image planes are identical and can be estimated from knowing that
the hard-edge mask has a diameter of 8 $\lambda / D$. A dashed line
in the `Lyot Plane' column indicates the outline of the circular
unobstructed entrance aperture. An $\sim$60\% throughput Lyot stop
was used for the hard-edge, Gaussian, and 4th-order BL masks. The
8th-order and 12th-order BL masks offer better rejection of
low-order aberrations at a cost of throughput and angular
resolution. The `Final Image' column shows the contrast generated by
each mask using the logarithmic scale in the hard-edge mask row; BL
masks remove on-axis starlight down to the numerical noise level of
the simulations ($< 10^{-12}$).} \label{tab:comparison}
\end{table}

\clearpage

\end{document}